# Geographically Weighted Canonical Correlation Analysis: Local Spatial Associations Between Two Sets of Variables

Zhenzhi Jiao[1], Angela Yao[1], Ran Tao[2*], Jean-Claude Thill[3]

*1.Urban Geospatial Analytics Lab, Department of Geography, University of Georgia, Athens, GA, USA*

*2.School of Remote Sensing Information Engineering, Wuhan University, Wuhan, China*

*3.Department of Earth, Environmental and Geographical Sciences, University of North Carolina at Charlotte, Charlotte, NC, USA*

*Correspondence: Ran Tao (rantao@whu.edu.cn)

**Abstract**

This article critically assesses the utility of the classical statistical technique of Canonical Correlation Analysis (CCA) to study spatial associations and proposes a new approach to enhance it. Unlike bivariate correlation analysis, which focuses on the relationship between two individual variables, CCA investigates associations between two sets of variables by finding pairs of linear combinations that are maximally correlated. CCA has great potential for uncovering complex multivariate relationships that vary across geographic space. We propose Geographically Weighted Canonical Correlation Analysis (GWCCA) as a new technique to explore local spatial associations between two sets of variables. GWCCA localizes standard CCA by weighting each observation according to its spatial distance from a target location, thereby estimating location-specific canonical correlations. GWCCA's effectiveness in recovering spatial structure and capturing spatial effects has been evaluated with synthetic data. A case study of US county-level health outcomes and social determinants of health is conducted to demonstrate GWCCA's capabilities empirically. It is concluded that GWCCA has the potential for a wide range of applications in spatial data–intensive fields such as urban planning, environmental science, public health, and transportation, where understanding local spatial associations in multivariate dimensions is crucial.

**Keywords**: Canonical Correlation Analysis; spatial association; multivariate; public health

## 1. Introduction

Multivariate analysis seeks to model and interpret the relationships among multiple interdependent variables by capturing their joint variation and underlying dependence structure, which enables inferences about shared latent factors, dimensionality, and cross-variable associations (Anselin, 2019; Chatfield, 2018; Lin, 2023; Mardia et al., 2024). Within this analytical tradition, Principal Component Analysis (PCA) and Factor Analysis (FA) have been widely employed to extract latent dimensions that account for the internal covariance structure of a multivariate dataset structured around a single semantic theme (Abdi & Williams, 2010; Gibson, 1959). Despite their effectiveness in summarizing intra-set dependence, these approaches are constrained in practice by the necessity of defining a theoretically sound, coherent and self-contained set of variables prior to analysis. Perhaps more fundamentally, however, many research questions reach beyond isolated effects to study relationships between two concepts, each linked to its own distinct set of variables that may capture separate but potentially interconnected dimensions. For instance, understanding how socioeconomic variables (e.g., income, educational attainment, employment status) relate to environmental indicators (e.g., air quality, green space accessibility, noise pollution) requires analyzing the interplay between two multivariate domains, which is beyond the one-on-one relationship of two variables or intra-set relationship of a set of variables. In such contexts, methods to capture the complex interdependencies between ensembles of variables are called for.

In contrast to techniques that characterize covariance within a single variable system, canonical ordination methods were developed to investigate associations between multiple interrelated variable systems (Borcard et al., 2018; Legendre & Legendre, 1998). This family of methods includes Canonical Correlation Analysis (CCA), Redundancy Analysis, and Canonical Correspondence Analysis, each suited to different data structures and analytical objectives. Among these methods, CCA focuses on identifying maximally correlated linear combinations between two sets of continuous variables, while Redundancy Analysis emphasizes explained variance through linear prediction, and Canonical Correspondence Analysis extends the approach to categorical or count data using chi-square distances. Focusing on shared cross-system covariance, CCA offers an exploratory framework for revealing structures of joint variation across multivariate domains (Hotelling, 1992; Yang et al., 2021). Its underlying principle entails finding pairs of linear combinations, known as canonical variates, in each variable set that maximize the correlation between them, thereby capturing the maximum shared multivariate structure across the domains. A distinctive feature of CCA is that it models the overall relationships between variable sets rather than a set of variables or individual variable pairs, making it uniquely suited for modeling complex interdependence among multiple variables (Härdle & Simar, 2015; Klatzky & Hodge, 1971).

Given these rich properties, CCA has been recognized for its great potential for uncovering geographic associations between two sets of variables (Gittins, 2012; Laessig & Duckett, 1979). Yet, CCA has been largely overlooked in the spatial data science domain, arguably due to its inability to explicitly accommodate the spatial structure of geographic data and to its interpretive complexity. Geospatial data inherently exhibit two fundamental properties: spatial autocorrelation, referring to the dependence among observations due to spatial proximity, and spatial heterogeneity, reflecting the spatial non-stationarity of statistical relationships or parameters across geographic space (Demšar et al., 2013; Fotheringham et al., 2002; Harris et al., 2014). Although methods such as Spatial Canonical Correlation Analysis (SCCA, Bhupatiraju et al. 2013) and Canonical Spatial Correlation Analysis (CSCA, Lin, 2025) have incorporated spatial autocorrelation into CCA calibration, their outputs have remained largely focused on global canonical correlations and loadings. As a result, they still fall short in providing location-specific analytical results and they remain limited in terms of spatial interpretability in local contexts.

Building on the above, it is appropriate to revisit the role of CCA in geographic research and to generalize it for spatial analysis. Traditional CCA assumes a homogeneous relationship among observations, producing global canonical correlation coefficients and variable loadings. More generally, however, spatial heterogeneity posits that relationships in spatial data is not stationary across locations, which poses challenges for interpretation and for the application of CCA in spatial analysis (A. S. Fotheringham & Brunsdon, 1999; Rogerson, 2008; Thill, 2011). Thus, explicitly incorporating spatial heterogeneity and developing a local version of CCA is essential to enhance its applicability and interpretability in geographic spatial analysis. For example, the nature and strength of the relationship between neighborhood socioeconomic status and multiple environmental exposures, such as air pollution, noise levels, and green space accessibility, may vary significantly across the urban space (Gray et al., 2013). A localized CCA model would allow for spatially adaptive estimation of these multivariate relationships, revealing place-specific patterns that traditional global CCA would overlook.

This study proposes Geographically Weighted Canonical Correlation Analysis (GWCCA) for localizing CCA to explicitly model spatial heterogeneity. GWCCA calibrates the CCA model by calculating the geographically weighted (GW) mean vector and GW variance-covariance matrix, allowing the canonical correlation coefficients and loadings to vary spatially. In particular, the development of GWCCA from CCA was inspired by the methodological shift from global to local models in consideration of spatial heterogeneity rather than spatial autocorrelation. The rest of this paper is organized as follows: Section 2 presents related methods and technical details of GWCCA. Section 3 validates the developed estimation procedures of GWCCA with synthetic data. Section 4 presents an empirical case study of relationships between health outcomes and social determinants of health using GWCCA. The last two sections discuss and conclude the study.

## 2. Background and methodology

Several classic spatial and statistical analysis techniques are particularly relevant to our new method. Some of them serve as the foundation, while others address similar analytical objectives. Figure 1 schematically illustrates these related concepts and methods.

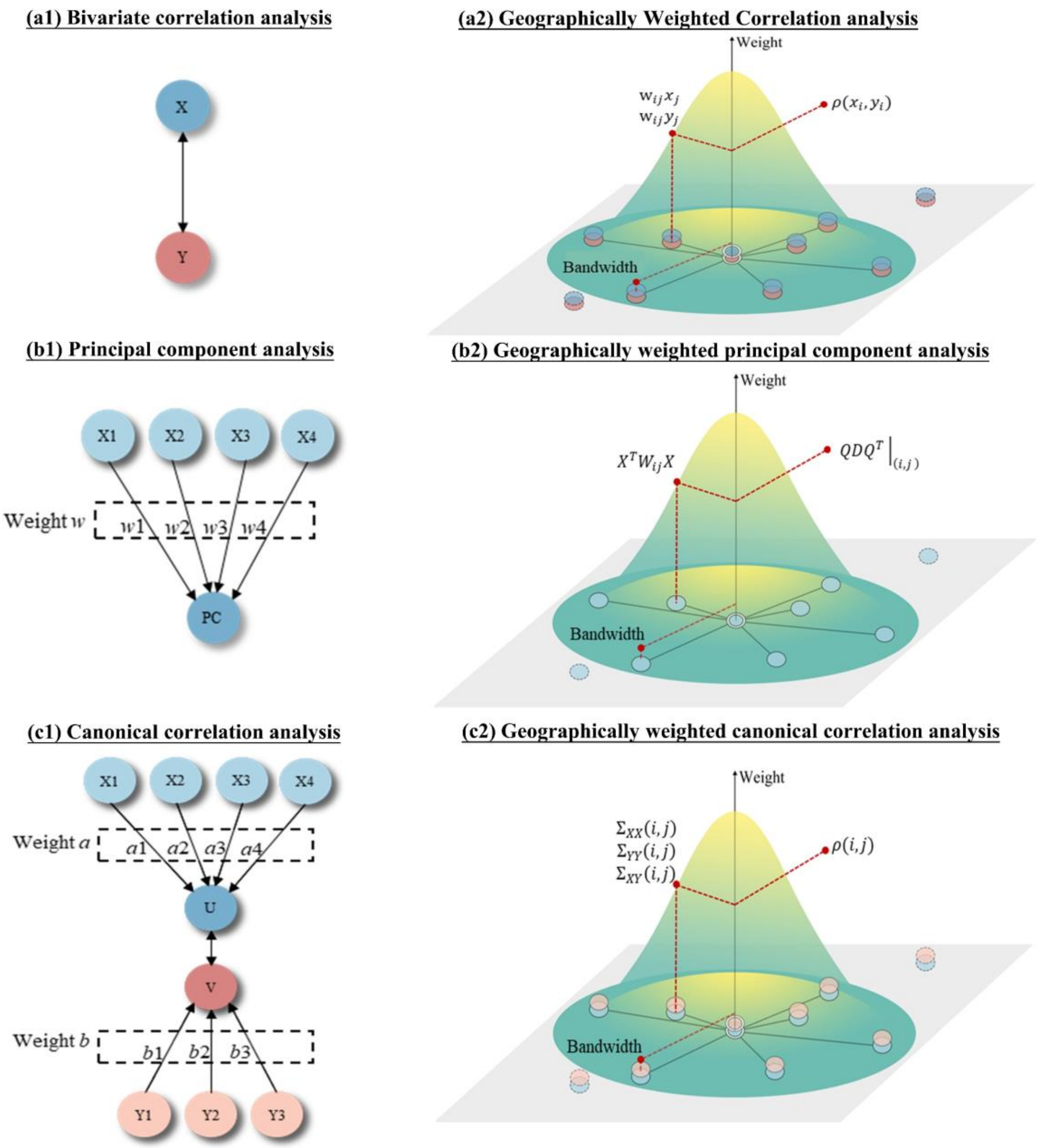

Figure 1. Relevant concepts and representative methods.

### *2.1 Bivariate Correlation Analysis and Geographically Weighted Correlation analysis*

As shown in Figure 1-a1, bivariate Correlation Analysis (BCA) is a statistical method to measure the correlation (degree and direction of association) between two variables (X and Y) (Haining, 1991; Wooditch et al., 2021). It is primarily used to investigate whether a linear relationship exists between two variables and to quantify the strength and direction of this relationship. Common methods include Pearson correlation coefficient, Spearman rank correlation, and Kendall correlation coefficient (Benesty et al., 2009; Valencia et al., 2019; Zar, 2005). However, BCA calculates the global correlation between two variables across the entire study area and cannot identify local spatial correlations (associations).

To capture the spatially varying association between two variables at a local level, Geographically Weighted Correlation Analysis (GWCA) was developed within the Geographically Weighted Summary Statistics (GWSS) and Geographically Weighted Local Statistics frameworks (Brunsdon et al., 2002; Fotheringham et al., 2002), with a unified and explicit formulation of the underlying GWSS measures later provided by Harris and Brunsdon (2010). GWCA evaluates the local correlation between variables at each spatial location by introducing a spatial weighting scheme, allowing correlations to adapt to local neighborhood structures and thereby revealing spatial heterogeneity in bivariate relationships. For each location $i$, the GW mean $m(x_i)$, standard deviation $s(x_i)$, and covariance $c(x_i, y_i)$, are computed as follows:

$$m(x_i) = \frac{\sum_{j=1}^{n} w_{ij} x_j}{\sum_{j=1}^{n} w_{ij}} \tag{1}$$

$$s(x_i) = \sqrt{\frac{\sum_{j=1}^{n} w_{ij} \left(x_j - m(x_i)\right)^2}{\sum_{j=1}^{n} w_{ij}}} \tag{2}$$

$$c(x_i, y_i) = \frac{\sum_{j=1}^{n} w_{ij} \left(x_j - m(x_i)\right)\left(y_j - m(y_i)\right)}{\sum_{j=1}^{n} w_{ij}} \tag{3}$$

The local correlation coefficient $\rho(x_i, y_i)$ at location $i$ is then defined as:

$$\rho(x_i, y_i) = \frac{c(x_i, y_i)}{s(x_i)s(y_i)} \quad (4)$$

### *2.2 Principal Component Analysis (PCA) and Geographically Weighted Principal Component Analysis*

As shown in Figure 1-b, PCA is a statistical technique of dimensionality reduction designed to preserve as much variance as possible in the data (Pearson, 1901). It transforms the original correlated variables with their weights ($w$) into a new set of uncorrelated variables called principal components (PC), which are ordered by the amount of variance they explain. Let $\boldsymbol{X} \in \mathbb{R}^{n \times m}$ denote the data matrix with $n$ observations and $m$ variables, , where $\boldsymbol{X}$ is assumed to be column-centered. The variance-covariance matrix $\boldsymbol{\Sigma} \in \mathbb{R}^{\mathrm{m} \times \mathrm{m}}$ is given by:

$$\boldsymbol{\Sigma} = \frac{1}{n-1}\boldsymbol{X}^T\boldsymbol{X} \quad (5)$$

where the diagonal elements represent the variances of each variable, and the off-diagonal elements represent the covariances between variables.

PCA is performed by solving the eigenvalue decomposition of the covariance matrix:

$$\boldsymbol{QDQ^T} = \boldsymbol{\Sigma} \quad (6)$$

where $\boldsymbol{Q}$ is an orthogonal matrix whose columns are the eigenvectors of $\boldsymbol{\Sigma}$ (i.e., the PC loadings), $\boldsymbol{D}$ is a diagonal matrix whose diagonal elements are the eigenvalues, representing the variance explained by each principal component.

The principal component scores are computed by projecting the observations onto the loading directions:

$$\boldsymbol{Z} = \boldsymbol{X}\boldsymbol{Q} \quad (7)$$

where each column of $\boldsymbol{Z}$ represents a principal component score.

Although PCA is highly effective in capturing the relationships within a set of variables, it only reveals global patterns and struggles to account for the inherent spatial heterogeneity in spatial data. To address this, Harris et al. (2011) proposed a local version of PCA for geospatial modeling, Geographically Weighted Principal Component Analysis (GWPCA). In the spatial domain, GWPCA calibrates the GW mean vector $\boldsymbol{\mu}(i, j)$, and the GW variance–covariance matrix $\boldsymbol{\Sigma}(i, j)$ to estimate the local principal components at geographic coordinates $(i, j)$:

$$\boldsymbol{\Sigma}(i,j) = \boldsymbol{X}^T \boldsymbol{W}_{ij} \boldsymbol{X} \quad (8)$$

where the local covariance structure is evaluated based on deviations from the geographically weighted local mean; $\boldsymbol{W}_{ij}$ is the diagonal matrix of geographic weights and is used as a kernel function in GWPCA. Typically, GWPCA uses a bi-square function:

$$w_{ij} = \begin{cases} \left[1 - \left(\frac{d_{ij}}{r}\right)^2\right]^2, & d_{ij} \le r \\ 0, & d_{ij} > r \end{cases} \quad (9)$$

where $d_{ij}$ is the distance between location $i$ and location $j$; $r$ is the bandwidth distance. GW principal components at location $(i, j)$ are obtained via the eigenvalue decomposition of the local covariance matrix as:

$$\boldsymbol{\Sigma}(i,j) = \boldsymbol{Q}(i,j)\boldsymbol{D}(i,j)\boldsymbol{Q}(i,j)^T \quad (10)$$

### *2.3 Canonical Correlation Analysis and Geographically Weighted Canonical Correlation Analysis*

Depicted in Figure 1-c1, CCA is a multivariate statistical method to uncover relationships between two sets of variables ($\boldsymbol{X}$ and $\boldsymbol{Y}$ set) by establishing and analyzing their linear combinations (weight vectors $\boldsymbol{a}$ and $\boldsymbol{b}$). This contrasts with BCA, which examines the correlation between two individual variables, and with PCA, which identifies relationships within a single set of variables (Hotelling, 1992; Laessig & Duckett, 1979). CCA posits that there are meaningful linear combinations of variables in each set and aims to find the combination of variables that are maximally correlated with each other.

Let $\boldsymbol{X} \in \mathbb{R}^{n \times p}$ and $\boldsymbol{Y} \in \mathbb{R}^{n \times q}$ be two sets of variables:

$$\boldsymbol{X} = (\boldsymbol{x}_1, \boldsymbol{x}_2, \dots, \boldsymbol{x}_p) \tag{11}$$

$$\boldsymbol{Y} = (\boldsymbol{y}_1, \boldsymbol{y}_2, \dots, \boldsymbol{y}_q) \tag{12}$$

where *n* is the number of observations or cases; *p* is the number of variables in set $\boldsymbol{X}$; *q* is the number of variables in set $\boldsymbol{Y}$.

CCA finds pairs of linear combinations, known as canonical variates, that maximize the correlation between two sets of variables, $\boldsymbol{X}$ and $\boldsymbol{Y}$. The correlation between each pair of canonical variates, referred to as the canonical correlation coefficient, ranges from 0 to 1. In CCA, the number of canonical variates is equal to the smaller number of variables in either of the $\boldsymbol{X}$ and $\boldsymbol{Y}$ sets. We seek weight vectors $\boldsymbol{a} \in R^p$ and $\boldsymbol{b} \in R^q$ such that the variables are transformed as follows:

$$U = \boldsymbol{X}\boldsymbol{a} \tag{13}$$

$$V = \boldsymbol{Yb} \tag{14}$$

where $\boldsymbol{a}$ and $\boldsymbol{b}$ are the weight vectors that maximize the correlation between the canonical variates $U$ and $V$. Specifically, the goal of CCA is to find linear transformations of $\boldsymbol{X}$and $\boldsymbol{Y}$that maximize the canonical correlation coefficient $\rho$ between them:

$$\rho = max \frac{\mathrm{Cov}(U, V)}{\sqrt{\mathrm{Var}(U)\mathrm{Var}(V)}} \tag{15}$$

Expanding the terms of covariance matrices:

$$\rho = \max_{\boldsymbol{a},\boldsymbol{b}} \frac{\boldsymbol{a}^T \boldsymbol{\Sigma}_{XY} \boldsymbol{b}}{\sqrt{\boldsymbol{a}^T \boldsymbol{\Sigma}_{XX} \boldsymbol{a} \cdot \boldsymbol{b}^T \Sigma_{YY} \boldsymbol{b}}} \tag{16}$$

where $\boldsymbol{\Sigma}_{XX}$ is the covariance matrix of $\boldsymbol{X}$; $\boldsymbol{\Sigma}_{YY}$ is the covariance matrix of $\boldsymbol{Y}$; $\boldsymbol{\Sigma}_{XY}$ is the cross-covariance matrix between $\boldsymbol{X}$ and $\boldsymbol{Y}$. This optimization problem can be solved with eigenvalue decomposition using a Lagrangian function:

$$\mathcal{L}(\boldsymbol{a}, \boldsymbol{b}, \lambda, \mu) = \boldsymbol{a}^T \boldsymbol{\Sigma}_{XY} \boldsymbol{b} - \lambda(\boldsymbol{a}^T \boldsymbol{\Sigma}_{XX} \boldsymbol{a} - 1) - \mu(\boldsymbol{b}^T \boldsymbol{\Sigma}_{YY} \boldsymbol{b} - 1) \tag{17}$$

By taking the derivatives with respect to $\boldsymbol{a}$ and $\boldsymbol{b}$, we obtain the following two generalized eigenvalue problems:

$$\boldsymbol{\Sigma}_{XX}^{-1} \boldsymbol{\Sigma}_{XY} \boldsymbol{\Sigma}_{YY}^{-1} \boldsymbol{\Sigma}_{YX} \boldsymbol{a} = \lambda \boldsymbol{a} \tag{18}$$

$$\boldsymbol{\Sigma}_{YY}^{-1} \boldsymbol{\Sigma}_{YX} \boldsymbol{\Sigma}_{XX}^{-1} \boldsymbol{\Sigma}_{XY} \boldsymbol{b} = \lambda \boldsymbol{b} \tag{19}$$

where $\lambda = \rho^2$ is the square of the canonical correlation coefficient; $\boldsymbol{a}$ and $\boldsymbol{b}$ are the canonical correlation vectors in the directions of $\boldsymbol{X}$and $\boldsymbol{Y}$, respectively.

Inspired by the extension from PCA to GWPCA (Harris et al., 2011), this study develops the GWCCA framework to locally estimate the GW mean vector and variance–covariance structure using a kernel weighting function. Like GWPCA, we assign weights to neighboring observations based on their distances to capture the spatially heterogeneous effects among the two sets of variables (Figure 1-c2). The spatial weight matrix $w_{ij}$ can be calculated by kernel functions (Koc & Akın, 2021). The GWCCA Python package[1] provides six kernel functions for users to choose, including Gaussian, Exponential, Box-car, Bi-square and Tri-cube (Gollini et al., 2015). For example, the Gaussian kernel function is given as:

$$w_{ij} = exp\left[-\frac{1}{2}\left(\frac{d_{ij}}{r}\right)^2\right] \tag{20}$$

where $d_{ij}$ is the distance between observation *i* and observation *j*; $r$ is the bandwidth distance, including fixed bandwidth and adaptive bandwidth; this study uses an adaptive bandwidth that sets the distance to the $k$-th nearest neighbor as bandwidth.

Using kernel functions, we obtain GW variance–covariance matrices:

$$\boldsymbol{\Sigma}_{\boldsymbol{XX}}(i,j) = \boldsymbol{X}^T \boldsymbol{W}_{ij} \boldsymbol{X} \tag{21}$$

$$\boldsymbol{\Sigma}_{\boldsymbol{YY}}(i,j) = \boldsymbol{Y}^T \boldsymbol{W}_{ij} \boldsymbol{Y} \tag{22}$$

$$\boldsymbol{\Sigma}_{\boldsymbol{XY}}(i,j) = \boldsymbol{X}^T \boldsymbol{W}_{ij} \boldsymbol{Y} \tag{23}$$

With GW variance–covariance matrices in hand, GWCCA proceeds by solving a locally weighted version of the standard CCA optimization problem. At each location $(i,j)$, GWCCA seeks local weight vectors $\boldsymbol{a}(\boldsymbol{i},\boldsymbol{j})$ and $\boldsymbol{b}(\boldsymbol{i},\boldsymbol{j})$ that maximize the local canonical correlation:

$$\rho(i,j) = \max_{a(i,j),\, b(i,j)} \frac{\boldsymbol{a(i,j)^T \Sigma_{XY}(i,j) b(i,j)}}{\sqrt{[\boldsymbol{a(i,j)^T \Sigma_{XX}(i,j) a(i,j)}][\boldsymbol{b(i,j)^T \Sigma_{YY}(i,j) b(i,j)}]}} \quad (24)$$

A crucial step in calibrating a GWCCA model lies in determining both the optimal $k$-th nearest neighbor's distance as bandwidth and the number of canonical variates ($c$) to retain for interpretation. The bandwidth governs the spatial scale of local estimation, while $c$ controls the dimensionality of the canonical space and the explanatory proportion of the cross-covariance structure between the two variable sets. For each candidate bandwidth $r$ and canonical variate $c$, we define a residual goodness-of-fit (RGOF) measure that quantifies the proportion of canonical correlation not explained by the first $c$ variates. Let $\rho_{ij}(r)$ denote the $j$-th local canonical correlation at spatial location $i$, and $\psi = \min(p,q)$ is the full canonical rank. The RGOF statistic is given by:

$$\text{RGOF}(c, r) = 1 - \frac{\sum_{i=1}^{n} \sum_{j=1}^{c} \rho_{ij}(r)^2}{\sum_{i=1}^{n} \sum_{j=1}^{\psi} \rho_{ij}(r)^2} \quad (25)$$

where $n$ is the number of observations. Smaller RGOF values indicate that the retained components capture a larger proportion of the total canonical association and, thus, represent a more compact, yet explanatory, model.

However, the bandwidth that minimizes the residual does not necessarily correspond to the true geographical context (Kwan, 2012; Spielman & Yoo, 2009). We suspect this is because the bandwidth optimization algorithm in GW models tends to prioritize numerical accuracy over spatial interpretability. This often results in overly large bandwidths that favor global trends, but suppress local variations and obscure fine-scale spatial detail. While bandwidth optimization can, in some contexts, also yield overly small bandwidths depending on the underlying spatial process and data characteristics, the focus here is to address the tendency toward excessively large bandwidths associated with over-smoothing. To mitigate this and reduce computational overhead, we implement an early stopping rule during the bandwidth scan: the process terminates when the relative improvement in RGOF falls below 1% for a specified number of consecutive steps. The default patience parameter is 2 and it typically ranges from 2 to 10, depending on the smoothness of the data and the desired sensitivity. This range was determined through repeated experiments.

The optimal number of canonical variates is determined by a two-step optimization strategy. In the first step, the spatial bandwidth was optimized jointly with a coarse screening of canonical variates to remove noise-dominated or spatially unstable components. This joint optimization alleviates over-smoothing and produces more robust local canonical patterns than optimizing alone. Conceptually, this first-step optimization follows the heuristic spirit of Tsutsumida et al. (2017)'s local criterion for GWPCA, which determines effective dimensionality using data-driven thresholds. Extending this idea to a correlation-based framework, we introduce a spatial stability criterion for GWCCA, retaining only those canonical variates whose coefficient supports remain salient across space. For each bandwidth $r$, candidate dimensions $\acute{c}$ are evaluated according to the spatial support of their coefficients. A canonical variate is regarded as spatially stable if its loadings are sufficiently salient across locations. Specifically, we define a magnitude threshold $\tau$ based on the empirical distribution of absolute loadings:

$$\tau = \phi(|\boldsymbol{A}| \cup |\boldsymbol{B}|) \tag{26}$$

where $\boldsymbol{A}$ and $\boldsymbol{B}$ denote the local canonical weight matrices for variables $X$ and $Y$, respectively. The operator $\phi(\cdot)$ returns the empirical quantile at level $\phi$, serving as a data-driven threshold. In this study, we set $\phi$=0.95, following common practice in adaptive thresholding and sparsity-based screening (Abramovich & Benjamini, 1996; Sun & Cai, 2009). To eliminate noise-driven or spatially unstable canonical variates, we apply a spatial stability screening procedure based on the support of canonical loadings. For each bandwidth, we evaluate candidate canonical variates using a support ratio $S_{c}(r)$, defined as the proportion of locations where a sufficient fraction of coefficients exceeds a magnitude threshold. Let $c_{im}^{(c)}$ be the $m$-th coefficient of the $c$-th canonical variate at location $i$, and К=$p$+$q$. The support ratio $S_{c}(r)$ is defined as the proportion of locations where at least a fraction $\alpha$ of coefficients exceed $\tau$:

$$S_{c}(r) = \frac{1}{n}\sum_{i=1}^{n} \mathbb{1}\left(\frac{1}{\text{К}}\sum_{m=1}^{d} \mathbb{1}\left(\left|c_{im}^{(c)}\right| > \tau\right) \geq \alpha\right) \tag{27}$$

The variability across is handled with a relative filtering rule that retains only those components whose support ratio exceeds a fraction $\beta$ of the mean support across all canonical dimensions:

$$S_{c}(r) \geq \beta \cdot \overline{S}(r) \quad \text{where} \quad \overline{S}(r) = \frac{1}{\psi}\sum_{j=1}^{\psi} S_j(r) \tag{28}$$

where $\psi$ is the full canonical rank. This spatial screening step helps eliminate noise-driven or locally unstable components, ensuring that retained variates reflect robust spatial patterns. It follows the principles of sparsity-based selection in high-dimensional inference (Fan & Lv, 2010; Meinshausen & Bühlmann, 2010), and provides a scale-free, interpretable criterion for determining the effective dimensionality at each bandwidth. In practice, $\beta$ is typically set between 0.7 and 0.9 to balance sensitivity and robustness; this study set =0.8.

In the second step, the number of meaningful canonical variates is determined according to the average strength of canonical correlations across space. In CCA, component-wise correlations above 0.30–0.40 are typically considered meaningful (Hair, 2009; Tabachnick & Fidell, 2019; Thompson, 1984). Following this convention, we conservatively set the reporting threshold to 0.40, so that only components with non-trivial explanatory strength are retained for interpretation.

## 3. Synthetic Data

### *3.1 Evaluation on synthetic data*

We design two synthetic datasets to evaluate GWCCA in a controlled environment. In a two-dimensional spatial domain, we denote a set of observations with coordinates $(i, j)$. For each observation, we simulate two random vectors ($X$, $Y$), where $X$ has $p$ variables and $Y$ has $q$ variables. To ensure that these random vectors contain spatially varying covariance structures, we build the covariance matrix as s function of the spatial coordinates $i$ and $j$. Spatially varying correlations are imposed on the first two pairs of canonical variates, while subsequent canonical structures, if present, exhibit slight spatial variation. This design ensures that the dominant source of spatial correlation is controlled and known, allowing us to directly assess the ability of GWCCA to accurately recover the target spatially varying canonical relationship.

In Dataset I (Figure 2), we generate 2,000 random spatial locations over the unit square. The first canonical variate increases linearly with the first spatial coordinate and is invariant to the second coordinate, whereas the second canonical variate follows a localized two-dimensional Gaussian pattern centered in the spatial domain. For Dataset

II (Figure 3), we generate a 60×60 regular grid of spatial locations and prescribe two spatially varying canonical variates with distinct structures. The first canonical variate is generated from a zero-mean Gaussian random field, resulting in smooth yet complex spatial variation with multiple local maxima. The second canonical variate follows a diagonal linear gradient across the spatial domain. Dataset I is designed to represent dominant large-scale spatial trends, whereas Dataset II incorporates a combination of irregular fine-scale variation and directional spatial gradients. Together, these datasets span a broad spectrum of spatial complexities commonly encountered in real-world spatial processes. Further details on the synthetic data design are provided in Appendix A.

We use RGOF to calculate the optimal bandwidth of GWCCA, with the two datasets adopting the 244-th and 116-th neighbors as their respective optimal bandwidths. Excluding canonical variate coefficients below 0.4, we retain only the first two canonical variates from both datasets. By visualizing the simulated $\rho(i, j)$ and the estimated $\hat{\rho}(i, j)$ through spatial coordinates $i$ and $j$ in the first two canonical variates, we can assess whether the GWCCA estimates are capable of recovering the inherent spatial effects in the synthetic datasets (computational time and resource of GWCCA are reported in Appendix D). Figure 2 presents both the simulated first canonical variate $\rho 1(i, j)$ and the second canonical variate $\rho 2(i, j)$, the estimated first canonical variate $\hat{\rho} 1(i, j)$ and the estimated second canonical variate $\hat{\rho} 2(i, j)$ in synthetic dataset I, and Figure 3 shows the results in synthetic dataset II, where it can be observed that four GWCCA estimated $\hat{\rho}(i, j)$ closely approximates the four true $\rho(i, j)$ in both datasets. Table 1 summarizes the quantitative comparison between GWCCA and CCA using mean absolute error (MAE) and root mean square error (RMSE). GWCCA is a local model that directly estimates spatially varying $\hat{\rho}(i, j)$, making it directly comparable with the ground truth field. In contrast, CCA produces only a single global canonical correlation, so their MAE and RMSE are computed by comparing that global value with the mean of the true $\rho(i,j)$. Compared with CCA, GWCCA reduces both MAE and RMSE by at least 50%, demonstrating a substantial improvement in accuracy and its strong ability to capture spatial heterogeneity.

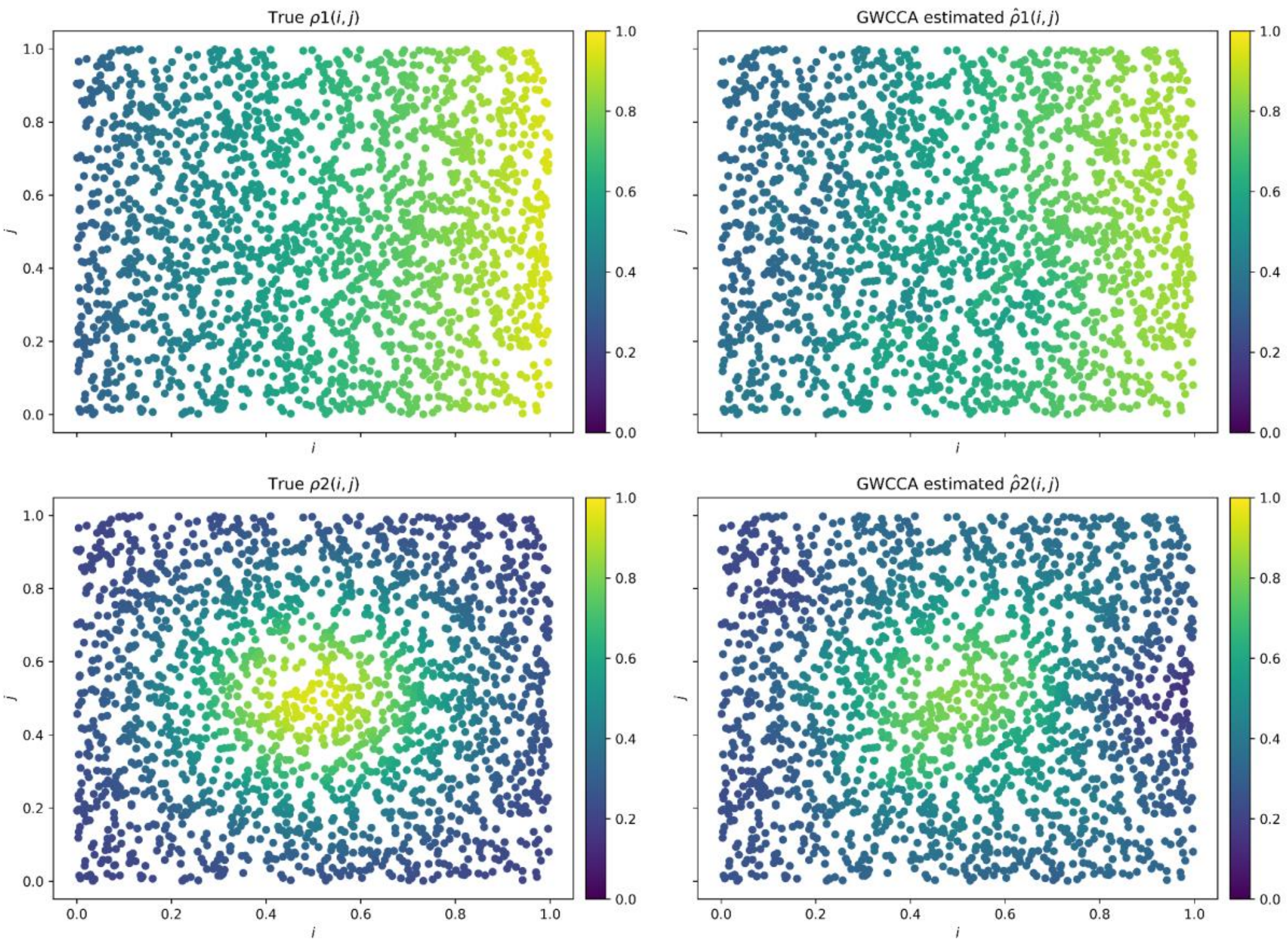

Figure 2. Spatial distribution of the first two canonical variates in Synthetic Dataset I and their GWCCA estimates.

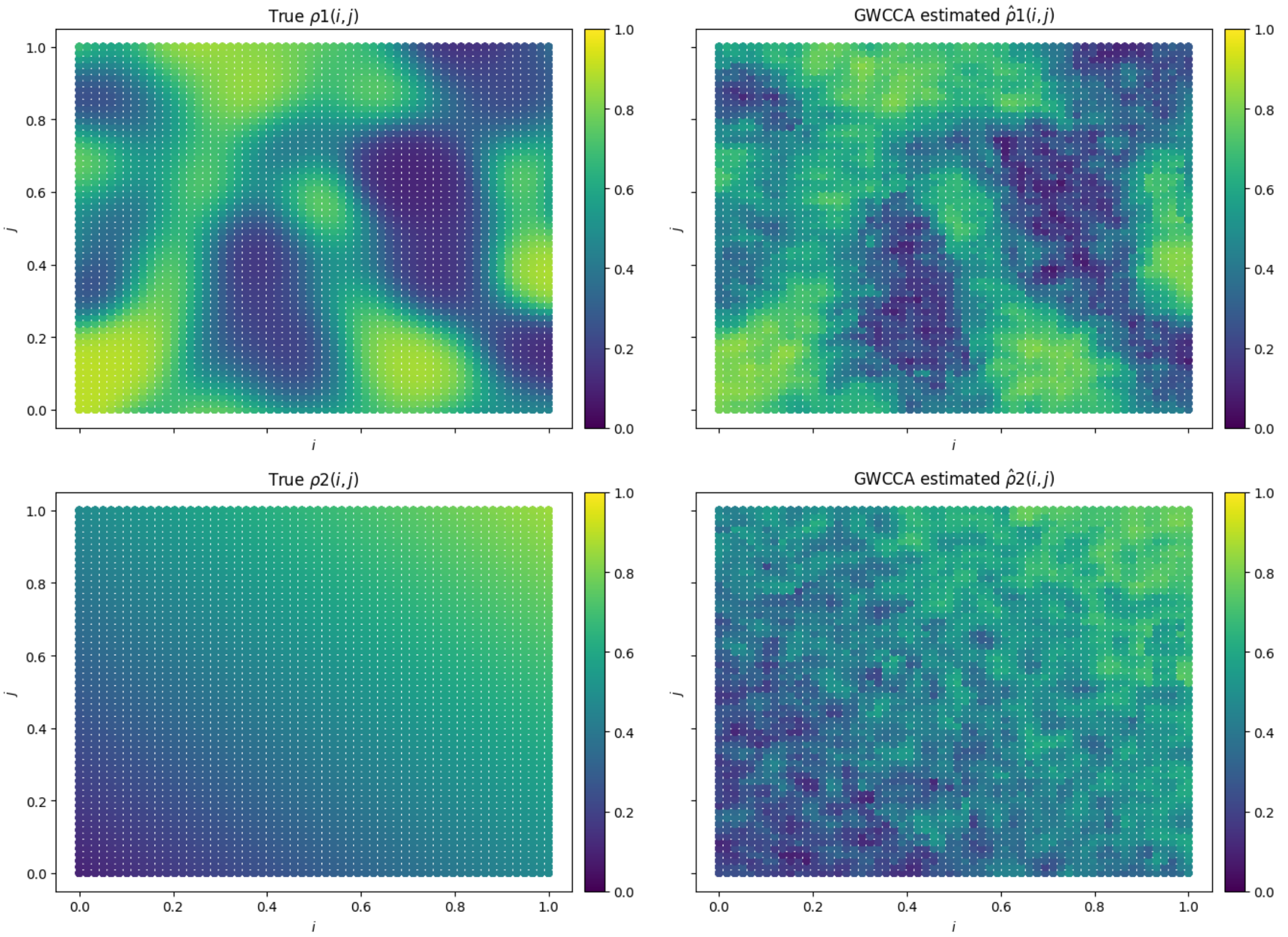

Figure 3. Spatial distribution of the first two canonical variates in Synthetic Dataset II and their GWCCA estimates.

Table 1. Model evaluations.

| | Canonical variates | Metrics | GWCCA | CCA |
|---|---|---|---|---|
| Synthetic dataset I | Canonical variate I | MAE | 0.0367 | 0.1625 |
| | | RMSE | 0.0453 | 0.1866 |
| | Canonical variate II | MAE | 0.0640 | 0.1413 |
| | | RMSE | 0.0768 | 0.1771 |
| Synthetic dataset II | Canonical variate I | MAE | 0.0611 | 0.1855 |
| | | RMSE | 0.0748 | 0.2145 |
| | Canonical variate II | MAE | 0.0621 | 0.1218 |
| | | RMSE | 0.0784 | 0.1508 |

### *3.2 Bandwidth comparations*

The early-stop RGOF builds on the idea that the regional goodness-of-fit initially rises with meaningful neighborhood expansion but plateaus once redundant information dominates, so the process stops automatically when further gain becomes negligible. To evaluate its performance, we compared three bandwidth selection schemes, RGOF without early stop, LOOR–CoV approach proposed by Tsutsumida et al. (2017), and early-stop RGOF—against the true $\rho(i, j)$. According to Table B1, Table B2, Figures 2-3, and Figures B1-B4 (see Appendix B), the GWCCA estimates based on the early stopping rule are far superior to those based on the LOOR–CoV approach, and in the vast majority of cases, they significantly outperform the RGOF method without early stopping. We observe that canonical variate II estimated using the RGOF method without early stopping outperforms the early stopping rule in terms of MAE in Synthetic dataset II. However, interestingly, further comparison reveals that the estimates obtained from the traditional RGOF method severely distort the underlying spatial relationships. This finding supports our earlier hypothesis that bandwidth optimization in GW models tends to prioritize numerical accuracy at the expense of spatial interpretability. The LOOR–CoV method improves numerical stability but still suppresses local variation and underestimates fine-scale contrasts. In contrast, the early-stop RGOF captures both broad regional gradients and local heterogeneity, yielding a correlation field that closely aligns with the true pattern while maintaining smooth spatial continuity. These results demonstrate that early-stop RGOF effectively alleviates the over-smoothing bias of conventional optimization, achieving a superior balance between spatial sensitivity, stability, and computational efficiency, and providing a more geographically meaningful criterion for bandwidth determination in GWCCA.

## 4. Case Study

Chronic diseases often exhibit co-occurrence, a phenomenon known as multimorbidity, which imposes a substantial burden on healthcare systems and has become a key concern in public health research (Pathirana & Jackson, 2018; Skou et al., 2022; Xu et al., 2024). Multimorbidity is shaped by the complex interplay of various social, economic, environmental, and racial factors (Alvarez-Galvez et al., 2023; Kolak, Bhatt, Park, Padrón, & Molefe, 2020). Previous studies have primarily focused on the global associations either among multiple diseases or between diseases and explanatory variables. However, in the context of spatial cross-sectional data, the presence of spatial heterogeneity renders global relationships fragile and potentially misleading.

### *4.1 Dataset and pre-processing*

Some studies have investigated the relationship between chronic diseases and social determinants of health (SDOH). Cockerham et al. (2017) identified SDOH as the fundamental causes of chronic health conditions. They discussed a wide range of chronic diseases (such as obesity, heart disease, cancer, cardiovascular disease, diabetes, stroke, pulmonary diseases, kidney disease, and others) and their associations with SDOH, including lifestyle, living and working environments, community characteristics, poverty, and environmental pollution. Xu et al. (2024) constructed a chronic disease network using nationwide county-level data and systematically identified comorbidity patterns among chronic diseases in the US, along with their associations with biological, psychological, socioeconomic, and behavioral factors. Kangas et al. (2025) conducted a systematic review examining the relationship between SDOH and health-related quality of life among patients with chronic conditions. Their findings indicated that age, gender, social support, marital status, education, employment, income, and urban residency were generally associated with health-related quality of life.

Based on the above literature on the relationship between chronic diseases (denoted as $X$) and SDOH (denoted as $Y$), this study identifies two sets of variables representing both domains. The 2020 county-level dataset (comprising 3,107 spatial units) of the contiguous US was collected from the Centers for Disease Control and Prevention (CDC). In CCA-type methods, removing highly correlated variables helps to avoid multicollinearity issues, improve computational stability, and ensure the interpretability of canonical variates (Härdle & Simar, 2015). Thus, we use Pearson correlation coefficients to filter out variables with a correlation exceeding 0.7 (Jiao & Tao, 2025a). The descriptive statistics of the retained variables are presented in Table 2.

We use z-score standardization to pre-process the dataset, ensuring both computational stability and interpretability. This transformation eliminates unit and scale differences, preventing dominant variables from biasing the analysis and stabilizing matrix computations. Consequently, the resulting canonical loadings and variates can be interpreted as standardized, comparable measures that clearly reflect the relative importance of each variable. This study uses early-stop RGOF criterion to determine the optimal bandwidth, which is the 100-th neighbor in the case study.

Table 2. Descriptive statistics of dataset.

| | Variables | Min | Mean | Max | std. | Description (%) |
|---|---|---|---|---|---|---|
| Chronic diseases | Arthritis | 15.8 | 26.02 | 36.5 | 3.66 | Percentage of people with arthritis |
| | Asthma | 7.4 | 10.47 | 14.1 | 1.01 | Percentage of people with asthma |
| | Cancer | 4.3 | 5.84 | 6.4 | 0.30 | Percentage of people with cancer |
| | Depression | 12.5 | 22.62 | 31.9 | 3.30 | Percentage of people with depression |
| | Stroke | 1.7 | 3.49 | 7.3 | 0.82 | Percentage of people with stroke |
| | Diabetes | 5.9 | 11.72 | 21.5 | 2.58 | Percentage of people with diabetes |
| Social determinants of health | Age65 | 3 | 18.86 | 45.3 | 4.373 | Percentage of people aged ⩾65 years |
| | White | 1 | 69.51 | 100 | 23.12 | Percentage of white population |
| | Binge | 8.2 | 16.52 | 27.1 | 3.03 | Percentage of adults who engage in binge drinking |
| | Poverty | 0 | 24.49 | 71 | 8.51 | Percentage of persons below poverty |
| | Black | 0 | 0.07 | 0.77 | 0.12 | Percentage of Black population |
| | Hispanic | 0.06 | 0.11 | 0.96 | 0.15 | Percentage of Hispanic population |

### *4.2 Results*

Five canonical variates of GWCCA and CCA are obtained in this study (computational time and resource of GWCCA are reported in Appendix D). The mean value of the first local canonical correlation coefficients is 0.981, followed by 0.897 for the second; the mean value of the third local canonical correlation coefficients is 0.620, while the fourth drops significantly to 0.341 (Table 3). In traditional global CCA, canonical variates with correlation coefficients greater than 0.4 are typically retained (Hair et al., 2010). In the same spirit, we recommend that, in GWCCA, the mean value of local canonical correlation coefficients should also exceed 0.4. In this paper, we focus on presenting the first two canonical variates with the highest canonical correlation coefficients, while summarizing the results for the third canonical variate in Appendix C.

Table 3. Summary statistics of canonical variates with Global CCA and GWCCA.

| | GWCCA | | | | | | CCA |
|---|---|---|---|---|---|---|---|
| | Min | 25% | 50% | 75% | Max | Mean | Overall |
| Variate 1 | 0.942 | 0.977 | 0.982 | 0.985 | 0.992 | 0.981 | 0.974 |
| Variate 2 | 0.736 | 0.871 | 0.908 | 0.928 | 0.967 | 0.897 | 0.888 |
| Variate 3 | 0.262 | 0.537 | 0.611 | 0.705 | 0.943 | 0.620 | 0.511 |
| Variate 4 | 0.108 | 0.257 | 0.325 | 0.399 | 0.683 | 0.341 | 0.291 |

In GWCCA, each local canonical variate can be understood as the local correlation between an index formed by a linear combination of variables from set $X$ and an index formed by a linear combination from set $Y$. In Figure 4, the first local canonical variate remains at a high level across the US, indicating a strong linear association between $X$ and $Y$ without sharp local deviation from the national pattern. This suggests that the first pairwise canonical variate captures the dominant mode of covariation between the two variable sets. In the Eastern US, particularly across the Southeast, Mid-Atlantic, and parts of the Midwest, the local first variates are relatively high, indicating a strong and spatially consistent covariance structure between $X$ and $Y$. This pattern likely reflects homogeneous demographic, health, or socioeconomic conditions and attitudes with respect to health issues that support stable multivariate associations. In contrast, significantly lower coefficients are observed in the Southwest, especially in New Mexico and western Texas, as well as parts of Nevada and Utah, suggesting distinct variable relationships in these regions compared to others. The second set of local canonical variates exhibit more pronounced non-stationarity across geographies, revealing clear spatial heterogeneity. The southern Central US and central Appalachian region, particularly including eastern Texas, southern Arkansas, northern Louisiana, West Virginia and eastern Kentucky, exhibit relatively high GWCCA coefficients for the second canonical variates, suggesting a robust linear relationship between $X$ and $Y$ in these areas along this second dimension. Meanwhile, significantly lower coefficients are observed across the Pacific Northwest (e.g., Washington, Oregon, and Idaho), as well as the Northern Plains (e.g., Montana, North Dakota).

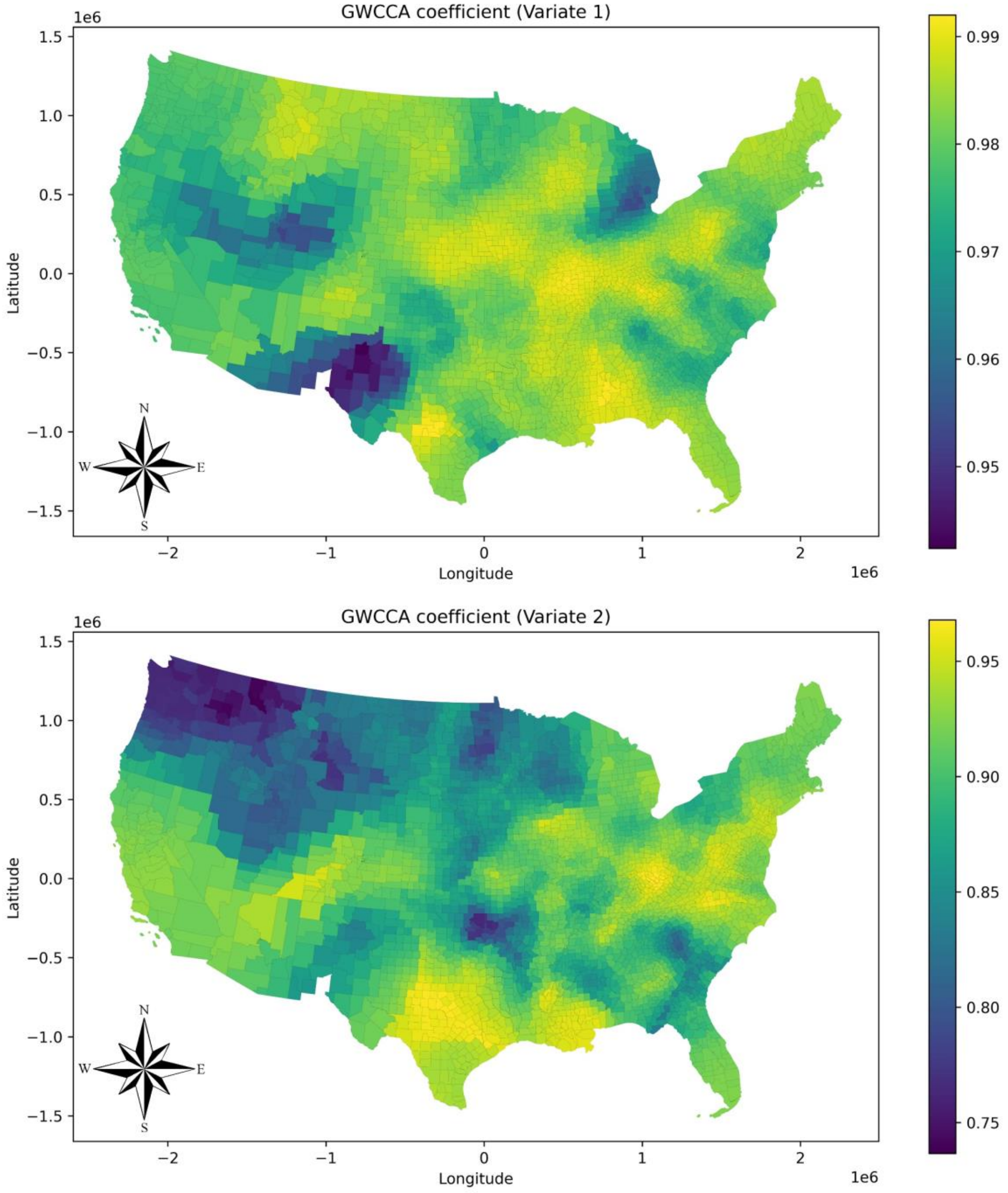

Figure 4. Local canonical variates in GWCCA.

Examining the variable loadings of these linear combinations is critically useful in interpreting the results of GWCCA in terms of relationships between health outcomes and SDOH. Tables 4 and 5 summarize the variable loadings. In particular, abs.mean refers to the average of the absolute values of all variable loadings for a variable, which indicates the influence of each original variable in its corresponding linear combination (Jiao & Tao, 2025b). Based on the information in these two tables, we can understand the meaning represented by the different linear combinations of each canonical variate, as well as the interpretation of the canonical variates themselves. For the first canonical variate, *White*, *Cancer*, and *Arthritis* load positively, whereas *Poverty*, *Stroke*, and *Diabetes* load negatively and show the greatest spatial variability. Consequently, the first canonical variate associates counties with a larger white population and lower poverty with higher arthritis and cancer prevalence, while most counties with higher poverty tend to display elevated stroke and diabetes rates. This represents the dominant covariation between health outcomes and SDOH across US counties.

As far as the second canonical variate is concerned, the mainly positive loadings are *White*, *Black*, *Arthritis*, *Stroke*, *Cancer* and *Poverty*; in contrast, *Diabetes*, *Binge* and *Hispanic* load negatively and display the widest spatial spread. This variate connects higher arthritis, stroke, and cancer rates to counties with larger, poorer white and Black populations. It brings into focus a pattern of association between health outcomes and SDOH that is at variance with that underscored by the first variate, namely higher diabetes, more binge drinking, and larger Hispanic populations. It also expresses a variant of associations observed in the first canonical variate, namely where arthritis, stroke and cancer co-occur with greater frequency in Black and White communities.

Table 4. Summary statistics of local variable loadings in local variate 1

| Variable | Min | 25% | 50% | 75% | Max | **Abs.mean** |
|---|---|---|---|---|---|---|
| Arthritis | 0.049 | 0.146 | 0.233 | 0.308 | 0.461 | **0.231** |
| Asthma | -0.401 | -0.177 | -0.086 | -0.016 | 0.101 | **0.117** |
| Cancer | -0.004 | 0.178 | 0.281 | 0.368 | 0.554 | **0.281** |
| Depression | -0.048 | 0.041 | 0.076 | 0.112 | 0.270 | **0.081** |
| Stroke | -0.885 | -0.710 | -0.625 | -0.521 | -0.014 | **0.596** |
| Diabetes | -0.854 | -0.641 | -0.523 | -0.422 | -0.093 | **0.525** |
| Age65 | -0.175 | -0.093 | -0.068 | -0.039 | 0.013 | **0.067** |
| White | 0.093 | 0.665 | 0.758 | 0.874 | 1.133 | **0.763** |
| Binge | -0.464 | -0.109 | -0.020 | 0.065 | 0.150 | **0.096** |
| Poverty | -0.904 | -0.722 | -0.666 | -0.574 | -0.062 | **0.650** |
| Black | -0.582 | -0.002 | 0.021 | 0.071 | 0.166 | **0.069** |
| Hispanic | -0.096 | 0.023 | 0.050 | 0.072 | 0.206 | **0.054** |

Table 5. Summary statistics of local variable loadings in local variate 2

| Variable | Min | 25% | 50% | 75% | Max | **Abs.mean** |
|---|---|---|---|---|---|---|
| Arthritis | 0.061 | 0.179 | 0.295 | 0.429 | 0.715 | **0.315** |
| Asthma | -0.327 | -0.036 | 0.048 | 0.145 | 0.428 | **0.117** |
| Cancer | -0.208 | 0.267 | 0.415 | 0.546 | 0.799 | **0.399** |
| Depression | -0.403 | -0.026 | 0.055 | 0.142 | 0.337 | **0.119** |
| Stroke | -0.509 | 0.312 | 0.476 | 0.589 | 0.763 | **0.445** |
| Diabetes | -0.741 | -0.427 | -0.306 | -0.145 | 0.412 | **0.307** |
| Age65 | -0.234 | -0.037 | -0.004 | 0.023 | 0.128 | **0.041** |
| White | -0.211 | 0.376 | 0.542 | 0.656 | 1.029 | **0.524** |
| Binge | -0.738 | -0.377 | -0.232 | -0.114 | 0.063 | **0.259** |
| Poverty | -0.437 | 0.279 | 0.369 | 0.473 | 0.695 | **0.372** |
| Black | -0.069 | 0.003 | 0.031 | 0.063 | 0.255 | **0.043** |
| Hispanic | -0.454 | -0.152 | -0.082 | -0.036 | 0.033 | **0.102** |

Given the previous findings, we start our analysis of geographic variations associated with the first canonical variate by focusing on the loadings for *Stroke*, *Diabetes*, *White*, and *Poverty*. In Figures 5 and 6, the negative *Stroke* loading is particularly prominent along the US East Coast and in the state of Washington. On the East Coast, this negative *Stroke* loading is more evident in areas characterized by lower levels of positive *White* loading and higher levels of negative *Poverty* loading. In southern Louisiana and western Texas, *Stroke* loading forms a low-value center, accompanied by similarly weak centers in both *White* loading and *Poverty* loading. Negative *Diabetes* loading is particularly prominent in the central US, showing a spatial pattern opposite to that of *White* loading. This contrast may suggest a stronger association with the distribution of minority populations. The spatial variation in *Poverty* loading is less pronounced in this region, indicating that poverty may not be strongly associated with diabetes in these areas.

Figures 7 and 8 display the local variable loadings of chronic diseases and social SDOH in the second canonical variate. Along the US East Coast, the linear combination of variables in *X* set is primarily characterized by positive loadings for *Stroke*, *Cancer*, and *Arthritis*, and a negative loading for *Diabetes*. In contrast, positive loadings for *White* and *Poverty* are particularly prominent in the linear combination of variables in *Y* set. This suggests a strong spatial association between the combination of *Stroke*, *Cancer*, and *Arthritis* and the combination of *White* and *Poverty* in this region. In the area surrounding California, a distinct geographic pattern emerges, where a negative linear combination of *Stroke* and *Depression* dominates the chronic disease loadings, while negative loadings of *Poverty* and *White* stand out in the SDOH loadings. This indicates a notable spatial association between *Stroke* and *Depression* with *Poverty* and *White* in the California region. In Texas, the linear combination of chronic diseases is mainly composed of positive loadings for *Arthritis*, *Diabetes*, *Asthma*, and *Stroke*, and a negative loading for *Depression*. These are closely associated with a combination of positive *Poverty* and *White* loadings, along with a negative loading for *Binge* drinking. These spatial associations highlight the importance for policymakers to consider how linear combinations of these variables vary across space in order to improve chronic disease management. To sum up, the case study of chronic diseases and SDOH across the US has demonstrated that GWCCA is effective at finding complex multivariate relationships between hypothesized concepts across the detailed county-level geographies evidenced through spatially-distributed data. This new method identifies core patterns of association and their variabilities that exist at different scales.

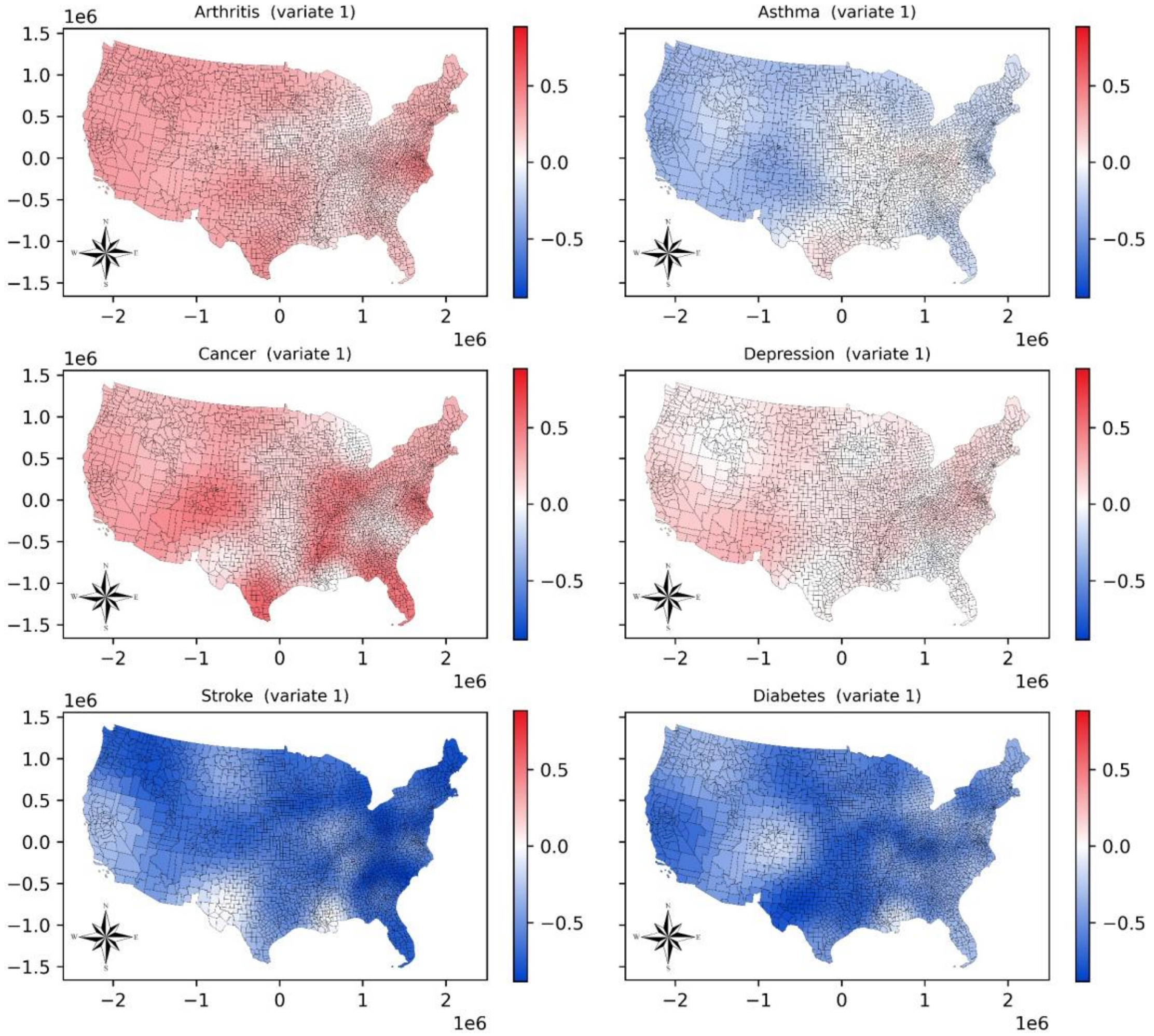

Figure 5. Local variable loadings of chronic diseases in canonical variate 1.

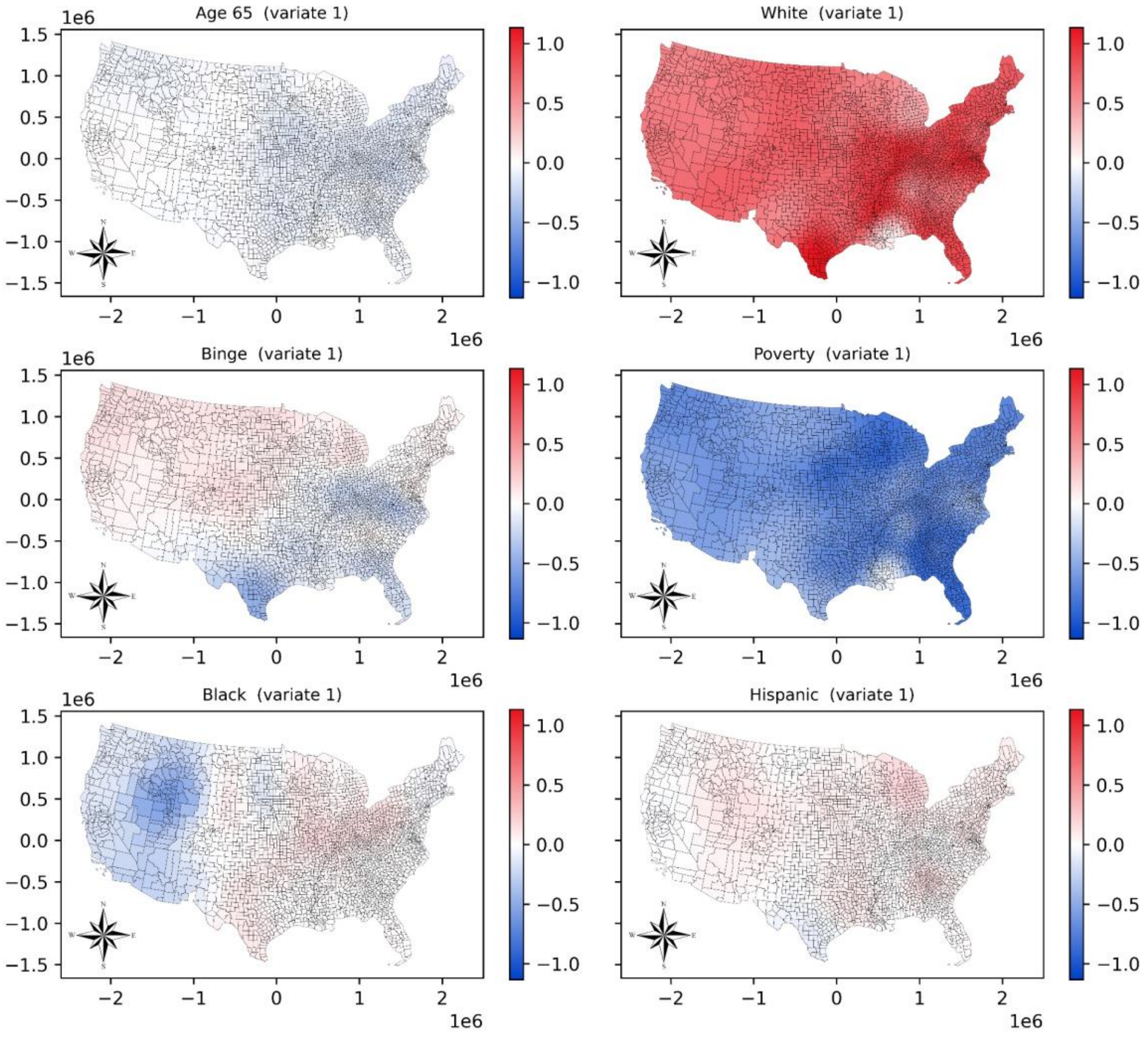

Figure 6. Local variable loadings of SDOH in canonical variate 1.

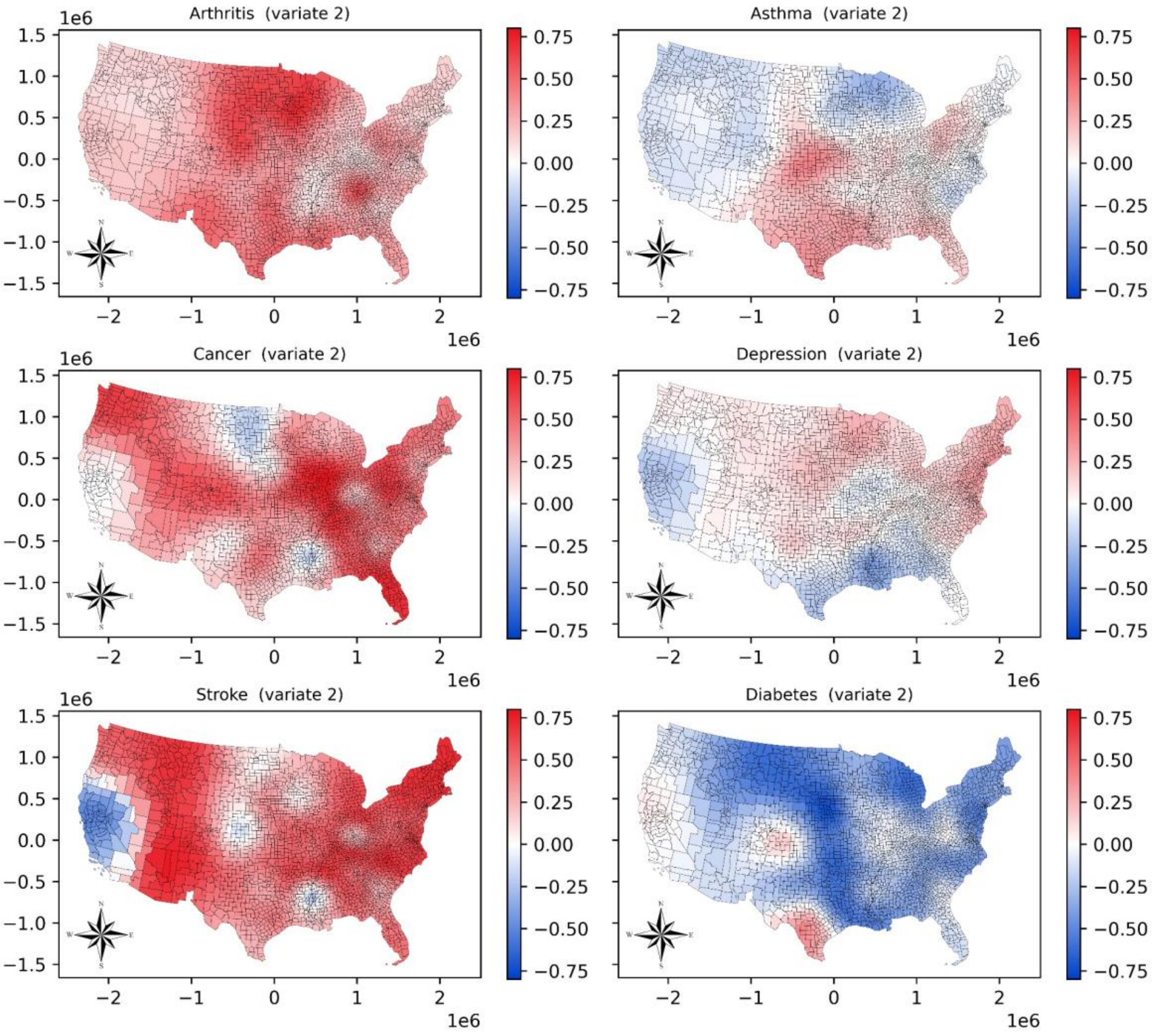

Figure 7. Local variable loadings of chronic diseases in canonical variate 2.

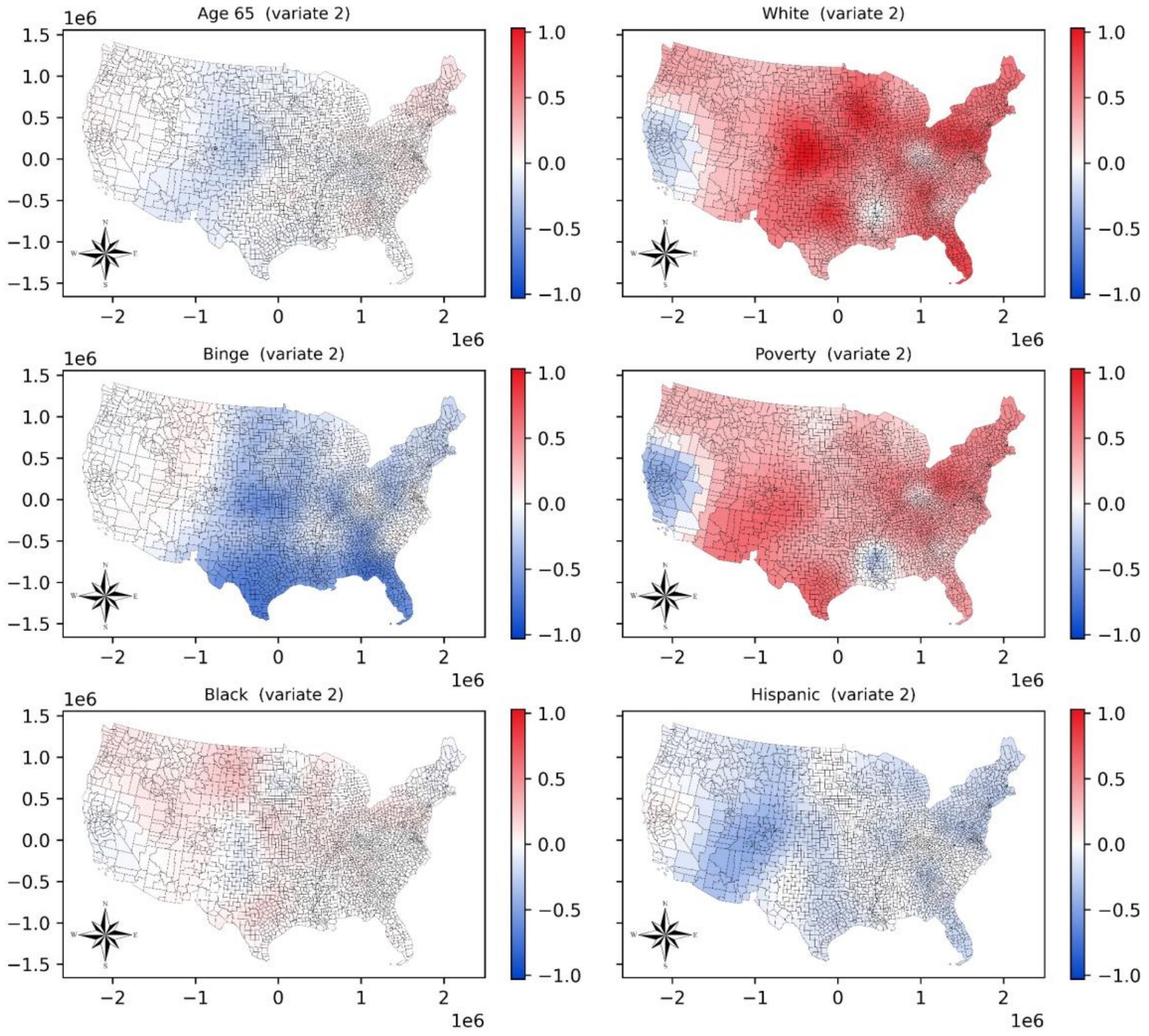

Figure 8. Local variable loadings of SDOH in canonical variate 2.

## 5. Discussion

Compared to other CCA-based models used to handle spatially distributed data and/or to address spatial problems, GWCCA offers several key advantages. First, existing spatial CCA methods, such as CSCA and SCCA, embed spatial structure in a global objective and also solve in one step, capturing only smooth, large-scale patterns. By contrast, GWCCA allows coefficients to vary across the geographic space in more complex ways by solving localized CCA problems, thereby uncovering fine-scale heterogeneity. More importantly, CSCA and SCCA lack built-in safeguards against Type I errors, relying on permutation tests and constrained model spaces to improve reliability. GWCCA, however, avoids confounding with spatial autocorrelation and supports resampling-based corrections, offering a more transparent and robust approach to false-positive control.

In GW models, the choice of bandwidth plays a decisive role in whether spatial heterogeneity can be properly captured. However, the optimization of bandwidth remains a subject of ongoing debate (Fotheringham et al., 2022; Koç, 2022). In this study, we argue that purely data-driven calibration tends to select excessively large bandwidths, which over-smooth spatial variation, obscure local patterns, and ultimately distort the underlying spatial structure. Our results corroborate these concerns and demonstrate that the proposed early-stop RGOF criterion offers a more balanced and effective solution. By monitoring the diminishing improvement trend in residual fit, early-stop RGOF adaptively identifies an appropriate spatial scale. The comparative experiments show that this approach effectively avoids over-smoothing, preserves fine-scale spatial details, and yields a more realistic reconstruction of spatial structures. At the same time, it is important to recognize that bandwidth selection is context dependent and is influenced by the scale of spatial variation, the spatial sampling configuration, and the stability of local estimation, such that optimization may yield values that are either excessively large or overly small. An early-stopping rule is appropriate for the former case and is generally unnecessary for the latter. Importantly, the early-stop RGOF framework is not limited to GWCCA; it can be extended to other GW models to enhance the robustness and interpretability of bandwidth optimization. Although our approach jointly optimizes the spatial bandwidth and the number of canonical variates, the final determination of retained variates still relies on a heuristic threshold adapted from conventional CCA practices. Future research should explore more principled, data-driven strategies to address this limitation.

In case study, the first canonical variate exhibits a relatively consistent canonical correlation coefficients across geographic space, indicating a stable pattern of association at the geographic scale, even though there is noticeable local variability in the loadings. In contrast, the other canonical variates demonstrate high sensitivity to spatial context in both variable loadings and canonical correlation coefficients strength. This contrast suggests that the first canonical variate captures the dominant layer of association between the two variable sets, while the other canonical variates reveal more nuanced, context-dependent spatial relationships that may be obscured in the primary mode. For instance, variables such as depression may exhibit spatial variation only in the second or third variate, as their effects could be masked by stronger health-related variables in the first canonical variate.

Unlike (GW) PCA, the signs of canonical loadings in GWCCA do not cause interpretational ambiguity. They simply indicate the direction of the canonical variates, and the interpretation focuses on the strength and consistency of co-variation between the two variable sets rather than on the sign itself. However, similar to the original CCA, GWCCA also faces issues of loading sign indeterminacy and rotational invariance (Gu & Wu, 2018; Makino, 2022). In GWCCA, due to the use of localized weighting, the sample characteristics within each region may vary greatly, leading to more random and unstable directions of local loading vectors. This uncertainty in local loadings makes cross-regional comparisons and interpretation more complex, as the same variable may exhibit opposite signs in different areas, thus increasing the challenge of interpretation. Another limitation of GWCCA is the absence of procedures for diagnosing omitted variable bias—a challenge that is prevalent across GW methods in general. Future work should aim to address this gap. In addition, GWCCA is suitable for exploring associations between two sets of variables. While it characterizes the relationship between two linear combinations, it is not designed for causal inference. Nevertheless, it remains a powerful tool for uncovering localized patterns of association in complex spatial datasets.

Future improvements to GWCCA can proceed in four directions. First, developing a nonlinear version of GWCCA would allow for the capture of potentially complex nonlinear relationships between the two sets of variables (Hsieh, 2000), thereby offering a more comprehensive understanding of their true interaction patterns. Second, extending GWCCA into a multi-scale framework would enable different variables to express their heterogeneity at varying spatial scales, like the concept behind Multiscale Geographically Weighted Regression (Fotheringham et al., 2017). Lastly, this paper emphasizes the methodological framework rather than the case study. Further research is needed to develop interpretative examples that illustrate the practical use of GWCCA.

## 6. Conclusions

We presented GWCCA, a local spatial statistical method, that incorporates a spatial weighting mechanism into canonical correlation analysis, allowing the canonical correlation coefficients to vary spatially. Compared to traditional spatial analysis methods, GWCCA offers unique advantages by explicitly examining the spatially varying relationships between two multivariate sets of variables, rather than the one-on-one relationship of two variables or the relationships within a set of variables. It integrates both exploratory and confirmatory aspects of spatial analysis. On the exploratory side, it identifies the combinations of variables that play a key role in the association between two sets of variables in different regions, thereby capturing structure of spatial associations and revealing spatial heterogeneity. On the confirmatory side, it verifies the significance of relationships and assesses the contribution of each variable through quantifying their loadings in canonical variates. The method can help explain the identified patterns of spatial associations at both global and local scales.

Because of its distinctive properties, GWCCA would be advantageously used across a broad range of disciplines such as ecology, environmental science, public health, and urban planning. In ecology, for example, GWCCA can serve to explore localized correlations between species distribution and environmental variables; in urban planning, GWCCA enables analysis of spatially heterogeneous covariation between crime types and socioeconomic as well as built environment factors, supporting more targeted policy interventions. This study also paves the way for extending GWCCA to other related techniques, such as canonical correspondence analysis, and presents its potential for modeling spatial flow interactions, an especially promising avenue for future research.

**Disclosure Statement**

No potential conflict of interest was reported by authors.

**Note**

1. GWCCA is released as an open-source Python package and is publicly available via PyPI (the Python Package Index). The package can be installed directly using the command pip install gwcca. The source code and examples are available at https://github.com/Josephjiao7/Geographically-Weighted-Canonical-Correlation-Analysis.

## Appendix A

### *Dataset I*

Let the spatial location be denoted by $(i, j)$. At each location $(i, j)$, the joint random vector $(\boldsymbol{X},\boldsymbol{Y}) \in R^{(p+q)\times(p+q)}$ follows a zero-mean multivariate Gaussian distribution with joint covariance matrix:

$$\boldsymbol{\Sigma}_{i,j} \in R^{(p+q)\times(p+q)} \tag{1}$$

The joint covariance matrix is specified as:

$$\boldsymbol{\Sigma}_{i,j} = \begin{pmatrix} \boldsymbol{I}_p & \boldsymbol{\Sigma}_{i,j}^{XY} \\ \boldsymbol{\Sigma}_{i,j}^{XY\top} & \boldsymbol{I}_q \end{pmatrix} + \varepsilon \boldsymbol{I}_{p+q} \tag{2}$$

where $\boldsymbol{I}_p$ and $\boldsymbol{I}_q$ are identity matrices, and $\varepsilon$>0 is a small jitter term added to ensure positive definiteness and numerical stability. The cross-covariance matrix between $\boldsymbol{X}$ and $\boldsymbol{Y}$ is constructed as:

$$\boldsymbol{\Sigma}_{i,j}^{XY} = \boldsymbol{A}_0 \begin{pmatrix} \rho_1(i,j) & 0 \\ 0 & \rho_2(i,j) \end{pmatrix} \boldsymbol{B}_0^\top \tag{3}$$

where $\boldsymbol{A_0} \in R^{p\times 2}$ and $\boldsymbol{B_0} \in R^{q\times 2}$ are fixed orthonormal matrices defining the global canonical directions for $\boldsymbol{X}$ and $\boldsymbol{Y}$, respectively. The diagonal matrix contains two spatially varying canonical correlations, $\rho_1(\mathrm{i,j})$ and $\rho_2(\mathrm{i,j})$, corresponding to the first and second canonical components. No additional canonical structures are present beyond these two components.

The two canonical correlations are specified to follow distinct spatial patterns. The first canonical correlation is defined as:

$$\rho_1(i,j) = 0.65\,i + 0.30 \tag{4}$$

which varies smoothly and monotonically along the $i$-direction. This component represents a dominant, large-scale spatial trend and constitutes the primary canonical mode in Dataset I.

The second canonical correlation is defined as:

$$\rho_2(i,j) = \beta_0 + \beta_1 \exp\left(-\frac{(i-c_1)^2 + (j-c_2)^2}{2\sigma^2}\right) \tag{5}$$

which forms a localized Gaussian bump centered at ($c1$, $c2$). This component introduces a weaker but spatially heterogeneous canonical relationship, capturing localized dependence patterns that are orthogonal to the large-scale trend represented by the first component. Both $\rho_1(i,j)$ and $\rho_2(i,j)$ are truncated below unity to ensure the positive definiteness of the joint covariance matrix.

To ensure identifiability and a well-defined canonical structure, the canonical directions for $\boldsymbol{X}$ and $\boldsymbol{Y}$ are assumed to be globally fixed and orthonormal across all spatial locations. Specifically, the matrices $\boldsymbol{A_0}$ and $\boldsymbol{B_0}$, which define the first two canonical directions of $\boldsymbol{X}$ and $\boldsymbol{Y}$, respectively, are composed of orthonormal columns.

$$\boldsymbol{A}_{\mathbf{0}}^{\top \boldsymbol{A}_0} = \boldsymbol{I}_2,\ \boldsymbol{B}_{\mathbf{0}}^{\top \boldsymbol{B}_0} = \boldsymbol{I}_{\mathbf{2}} \tag{6}$$

Under the above construction, each spatial location $(\mathrm{i}, \mathrm{j})$ is associated with a local joint distribution. Conditional on the spatial location $(\mathrm{i}, \mathrm{j})$, the random vector $(\boldsymbol{X}, \boldsymbol{Y})$ follows a zero-mean multivariate normal distribution, with its dependence structure fully characterized by the joint covariance matrix $\boldsymbol{\Sigma}_{\boldsymbol{i},\boldsymbol{j}}$.

$$(\boldsymbol{X}, \boldsymbol{Y}) \mid (i, j) \sim \mathcal{N}\left(\mathbf{0}, \boldsymbol{\Sigma}_{\boldsymbol{i},\boldsymbol{j}}\right) \tag{7}$$

***Dataset II***

Let $(i, j) = [0,1]^2$ denote the spatial domain, discretized into a regular grid of $n = s \times s$. We first simulate a zero-mean Gaussian random field (GRF) $Z(i, j)$ with a squared exponential covariance kernel:

$$\mathrm{Cov}\big(Z(i, j), Z(i', j')\big) = \sigma^2 \exp\left(-\frac{|(i, j) - (i', j')|^2}{2l^2}\right) \tag{8}$$

where *l* is the spatial scale (bandwidth), and $\sigma$ is the marginal variance.

The raw GRF $Z(i, j)$ is transformed into the canonical correlation field using a bounded non-linear transformation:

$$\rho_1(i, j) = 0.5 + 0.4 \cdot \tanh\left(\alpha \cdot \frac{Z(i, j) - E[Z]}{\mathrm{Std}(Z)}\right) \tag{9}$$

This guarantees that $\rho_1(i,j) \in (0.1, 0.95)$ and retains spatial smoothness inherited from the GRF kernel. This construction preserves spatial smoothness while introducing nonlinear variation in the first canonical correlation.

To introduce a secondary canonical structure with a simpler and more interpretable spatial pattern, the second canonical correlation is defined as a deterministic diagonal gradient:

$$\rho_2(i,j) = \rho_{\text{base}} + \rho_{\text{amp}} \frac{i+j}{2} \tag{10}$$

where $\rho_{\text{base}}$ and $\rho_{\text{amp}}$ control the baseline level and gradient strength, respectively.

Let $\boldsymbol{A_0} \in R^{p\times 2}$ and $\boldsymbol{B_0} \in R^{q\times 2}$ denote fixed orthonormal matrices defining the global canonical directions for $\boldsymbol{X}$ and $\boldsymbol{Y}$. These directions are shared across all spatial locations and satisfy:

$$\boldsymbol{A}_\mathbf{0}^\top \boldsymbol{A}_\mathbf{0} = \boldsymbol{I}_\mathbf{2},\ \boldsymbol{B}_\mathbf{0}^\top \boldsymbol{B}_\mathbf{0} = \boldsymbol{I}_\mathbf{2} \tag{11}$$

At each spatial location $(i,j)$, the cross-covariance matrix between $X(i,j)$ and $Y(i,j)$ is constructed as:

$$\boldsymbol{\Sigma}_{i,j}^{XY} = \boldsymbol{A}_\mathbf{0} \begin{pmatrix} \rho_1(i,j) & 0 \\ 0 & \rho_2(i,j) \end{pmatrix} \boldsymbol{B}_\mathbf{0}^\top \tag{12}$$

which induces two canonical components with spatially varying strengths but fixed directions.

The full joint covariance matrix at location $(i,j)$ is given by

$$\boldsymbol{\Sigma}_{i,j} = \begin{pmatrix} \boldsymbol{I}_\mathbf{p} & \boldsymbol{\Sigma}_{i,j}^{XY} \\ \boldsymbol{\Sigma}_{i,j}^{XY\top} & \boldsymbol{I}_\mathbf{q} \end{pmatrix} + \varepsilon \boldsymbol{I}_\mathbf{p+q} \tag{13}$$

where $\varepsilon>0$ is a small jitter term added for numerical stability.

Accordingly, the joint distribution of $\boldsymbol{X}(i,j)$and $\boldsymbol{Y}(i,j)$ is:

$$(\boldsymbol{X},\boldsymbol{Y}) \mid (i,j) \sim \mathcal{N}\left(0,\boldsymbol{\Sigma}_{i,j}\right) \tag{14}$$

which ensures that $\boldsymbol{X}(i,j) \sim \mathcal{N}\left(0,\boldsymbol{I}_p\right)$, $\boldsymbol{Y}(i,j) \sim \mathcal{N}\left(0,\boldsymbol{I}_q\right)$, and that the first and second canonical correlations equal $\rho_1(i,j)$ and $\rho_2(i,j)$, respectively.

**Appendix B**

Table B1. Model evaluations.

| | Canonical variates | Metrics | Early-stop RGOF | RGOF | LOOR-CoV |
|---|---|---|---|---|---|
| Synthetic dataset I | Canonical variate I | MAE | 0.0367 | 0.0605 | 0.0496 |
| | | RMSE | 0.0453 | 0.0775 | 0.0644 |
| | Canonical variate II | MAE | 0.0640 | 0.1375 | 0.1207 |
| | | RMSE | 0.0768 | 0.1577 | 0.1383 |
| Synthetic dataset II | Canonical variate I | MAE | 0.0611 | 0.1626 | 0.1649 |
| | | RMSE | 0.0748 | 0.1911 | 0.2034 |
| | Canonical variate II | MAE | 0.0621 | 0.0595 | 0.1204 |
| | | RMSE | 0.0784 | 0.0823 | 0.1506 |

Table B2. Bandwidth comparations (bandwidth measured as the number of k-nearest neighbors).

| | Early-stop RGOF | RGOF | LOOR-CoV |
|---|---|---|---|
| Synthetic dataset I | 244 | 995 | 795 |
| Synthetic dataset II | 116 | 1500 | 30 |

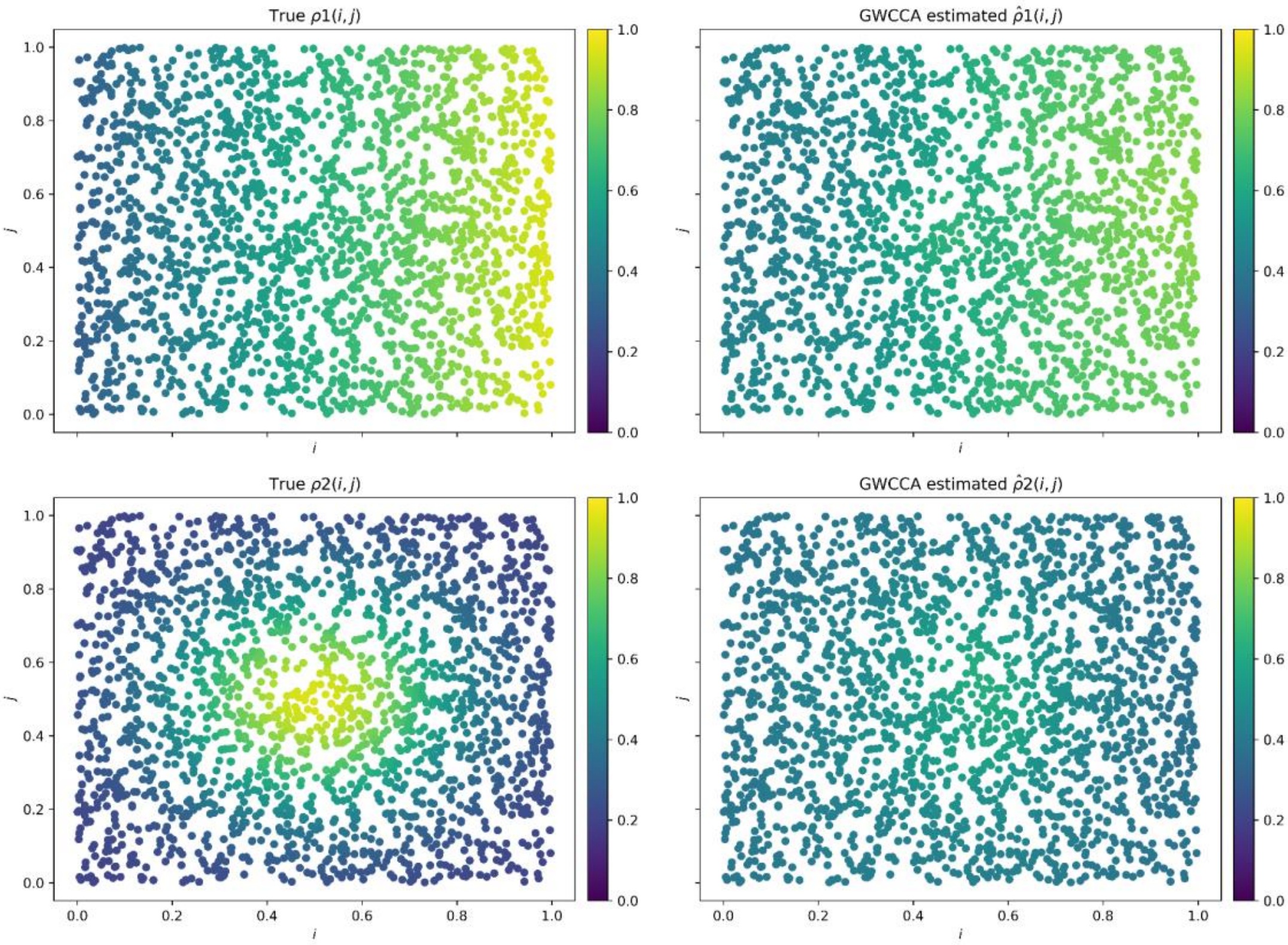

Figure B1. Illustration of LOOR–CoV bandwidth selection in Synthetic Dataset I

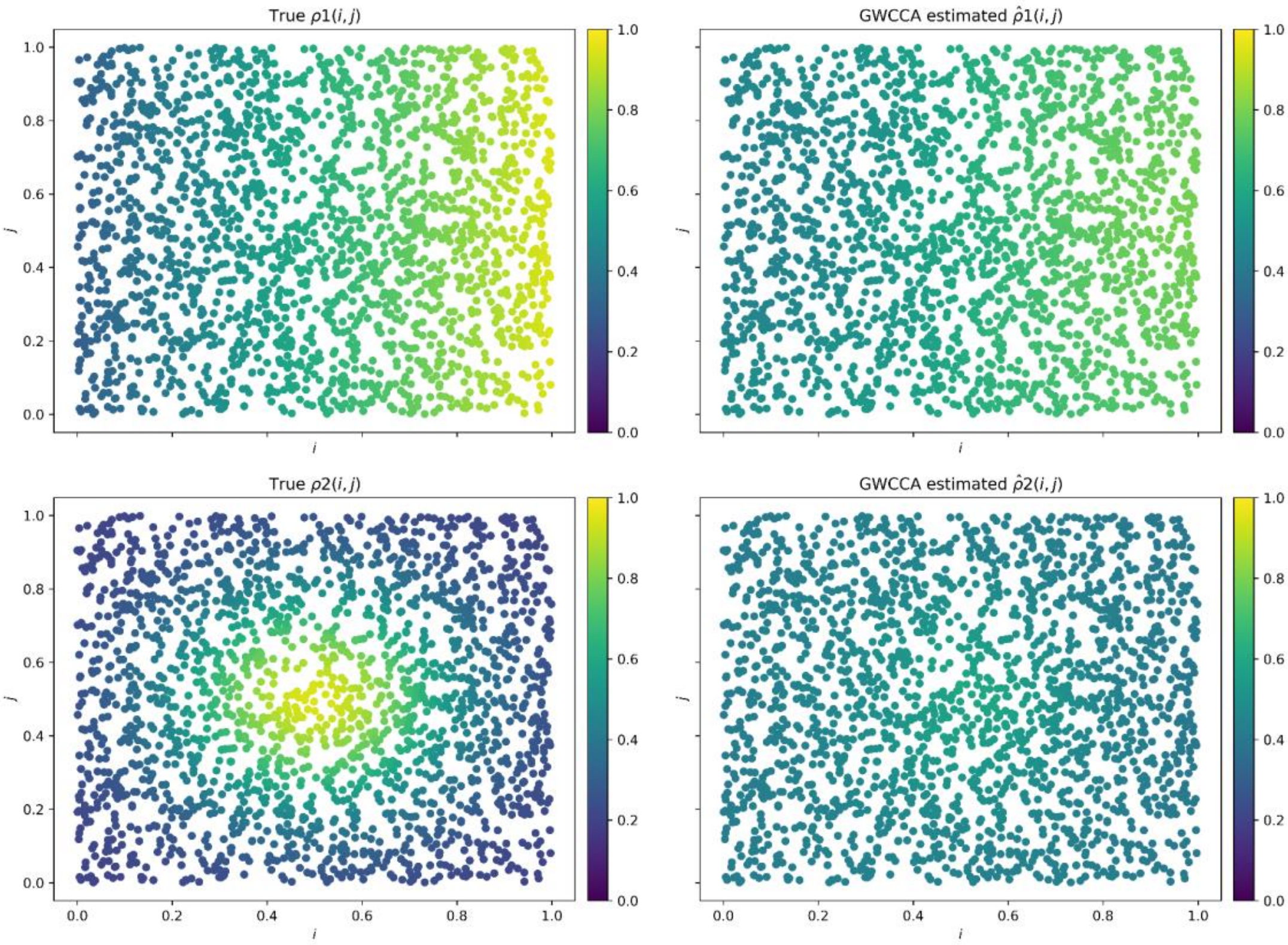

Figure B2. Illustration of bandwidth selection using RGOF in Synthetic Dataset I

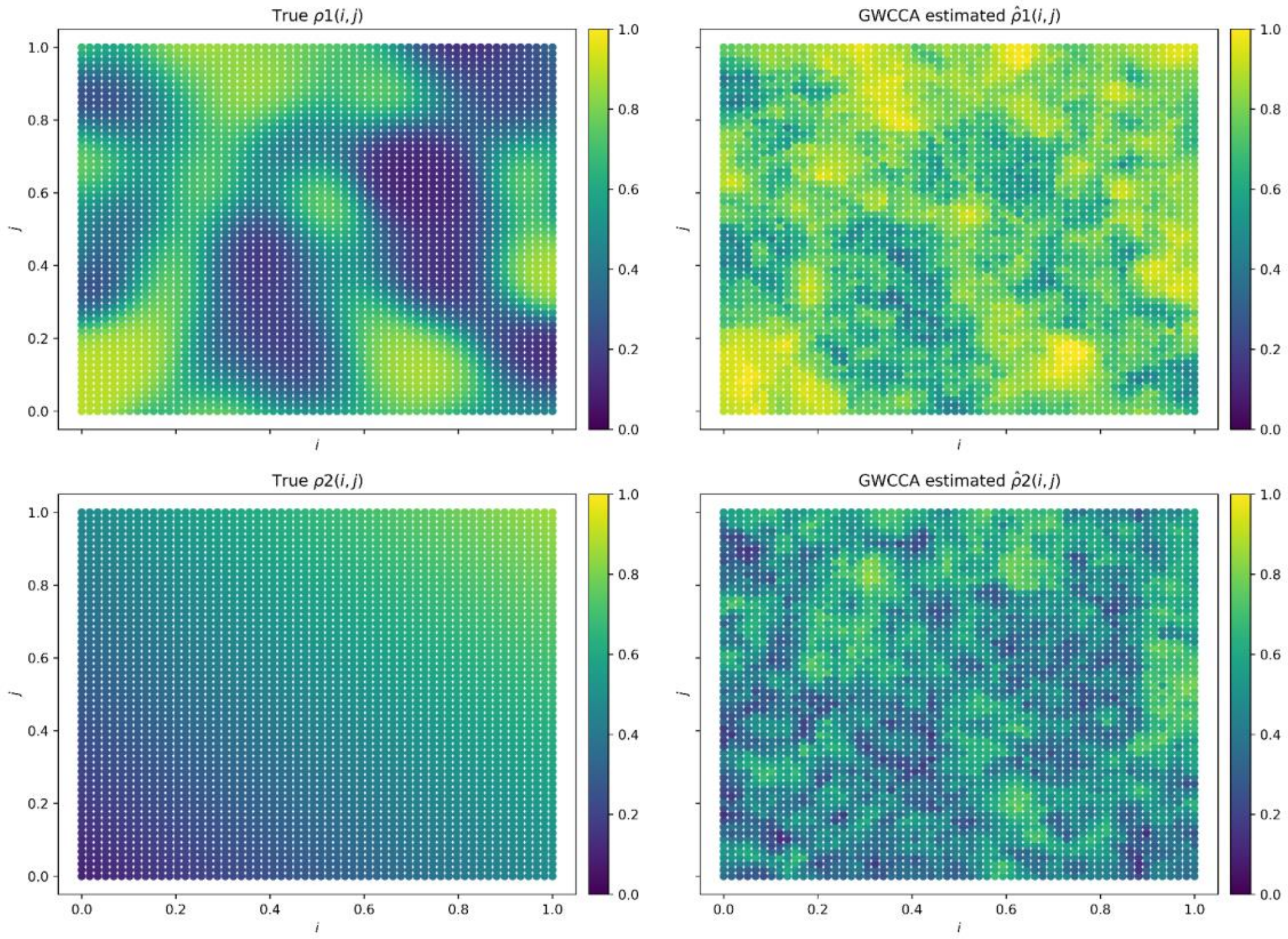

Figure B3. Illustration of LOOR–CoV bandwidth selection in Synthetic Dataset II

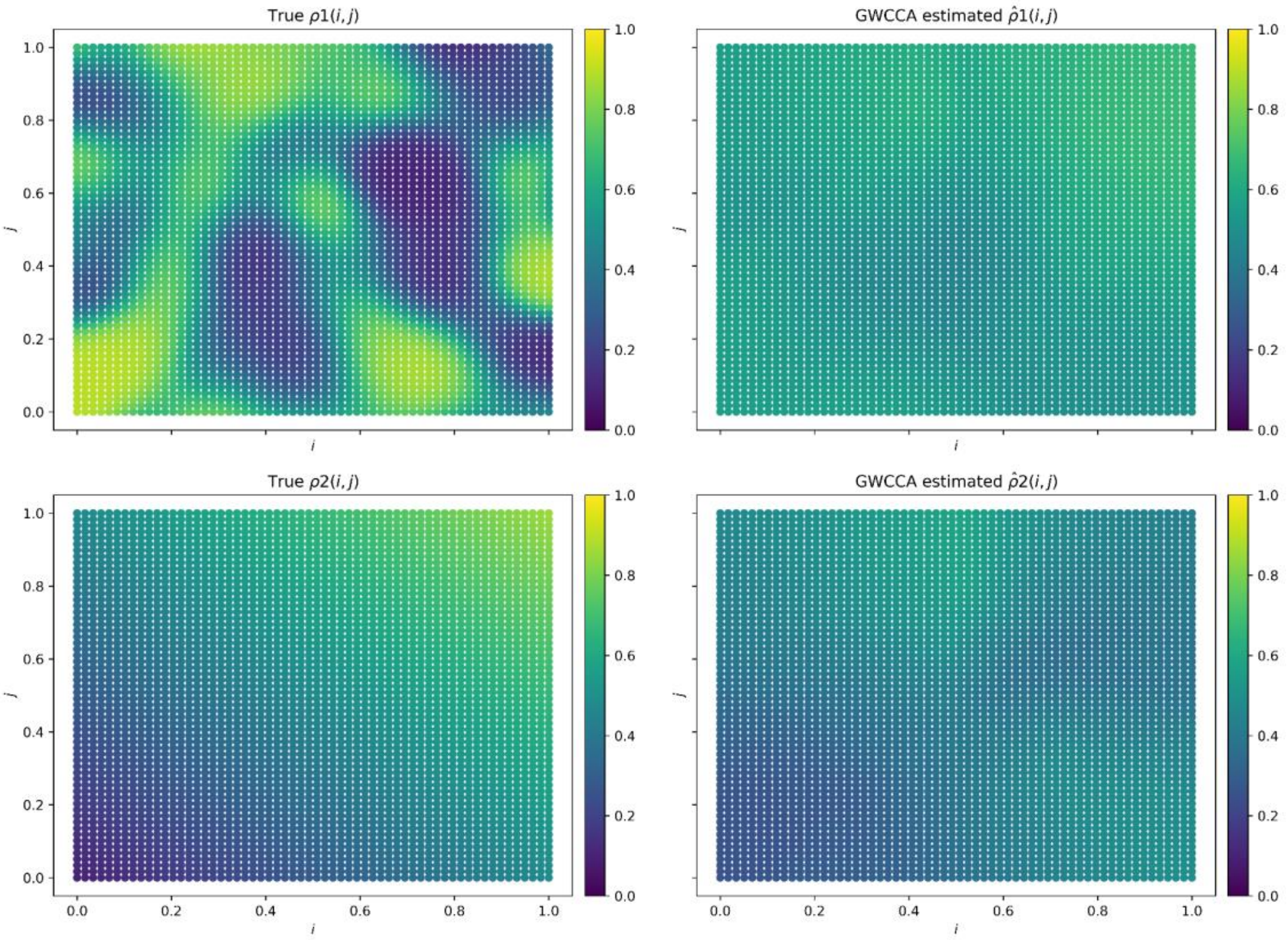

Figure B4. Illustration of bandwidth selection using RGOF in Synthetic Dataset Ⅱ

**Appendix C**

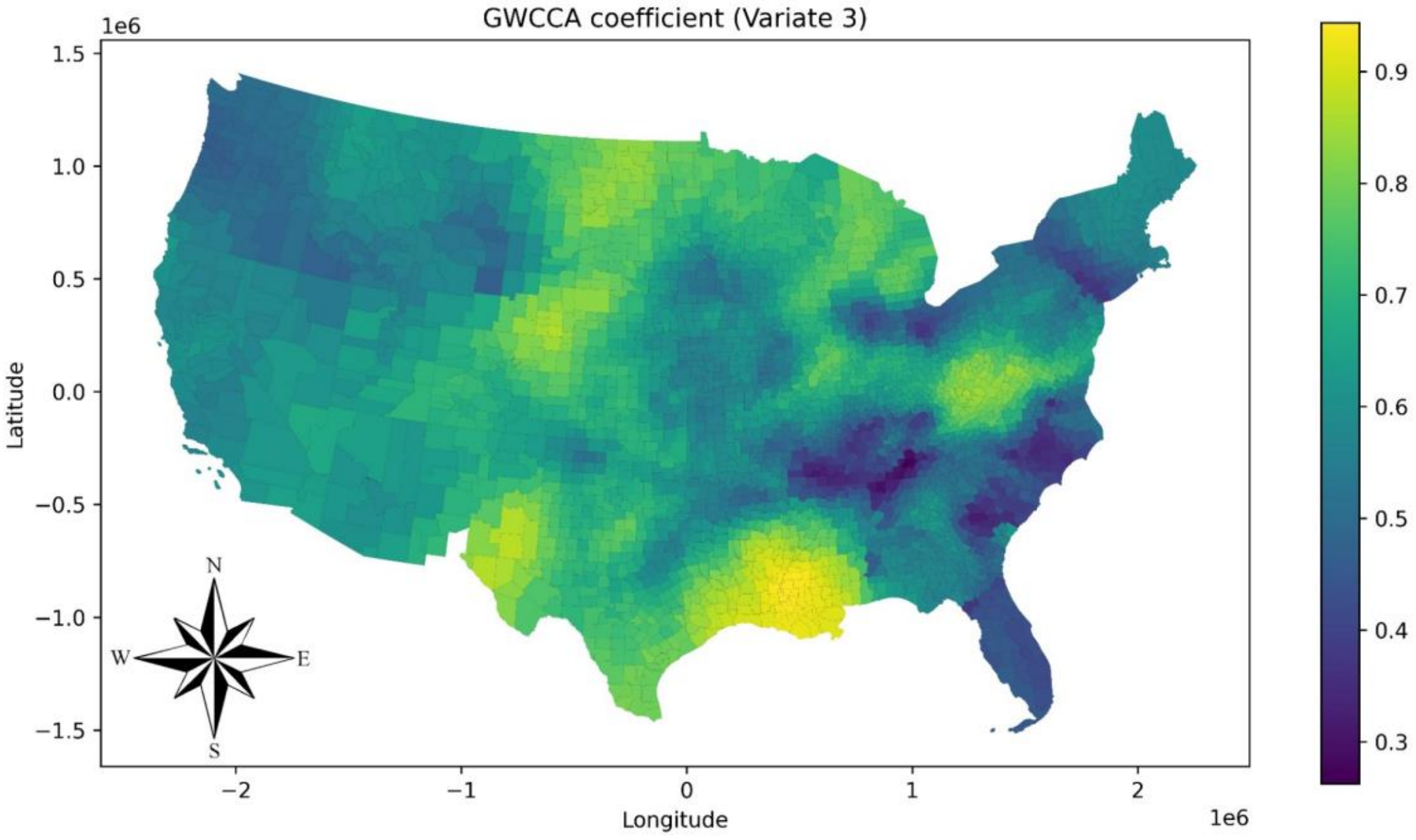

Figure C1. The spatial distribution of the third canonical variate

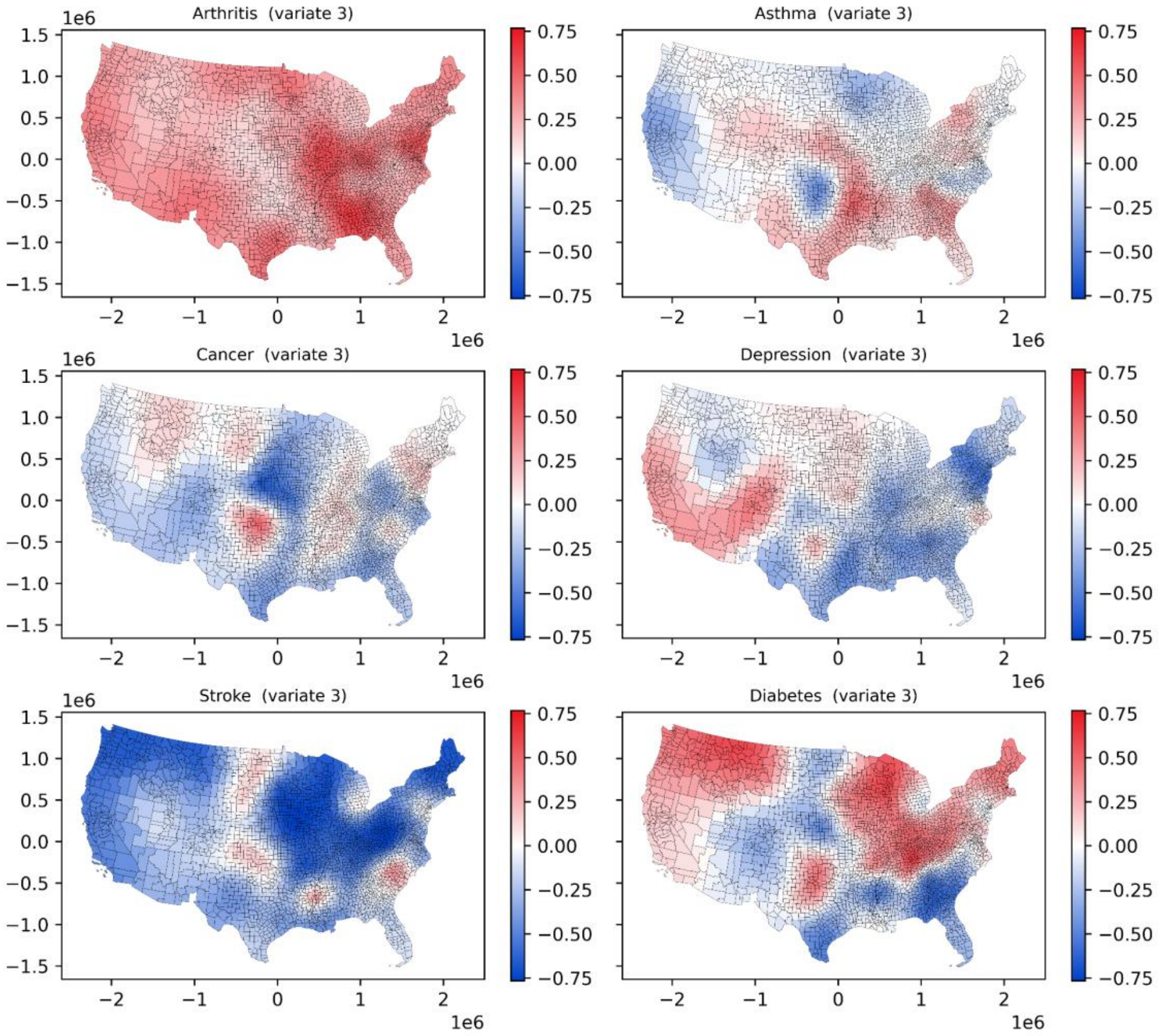

Figure C2. Local variable loadings of chronic diseases in canonical variate 3

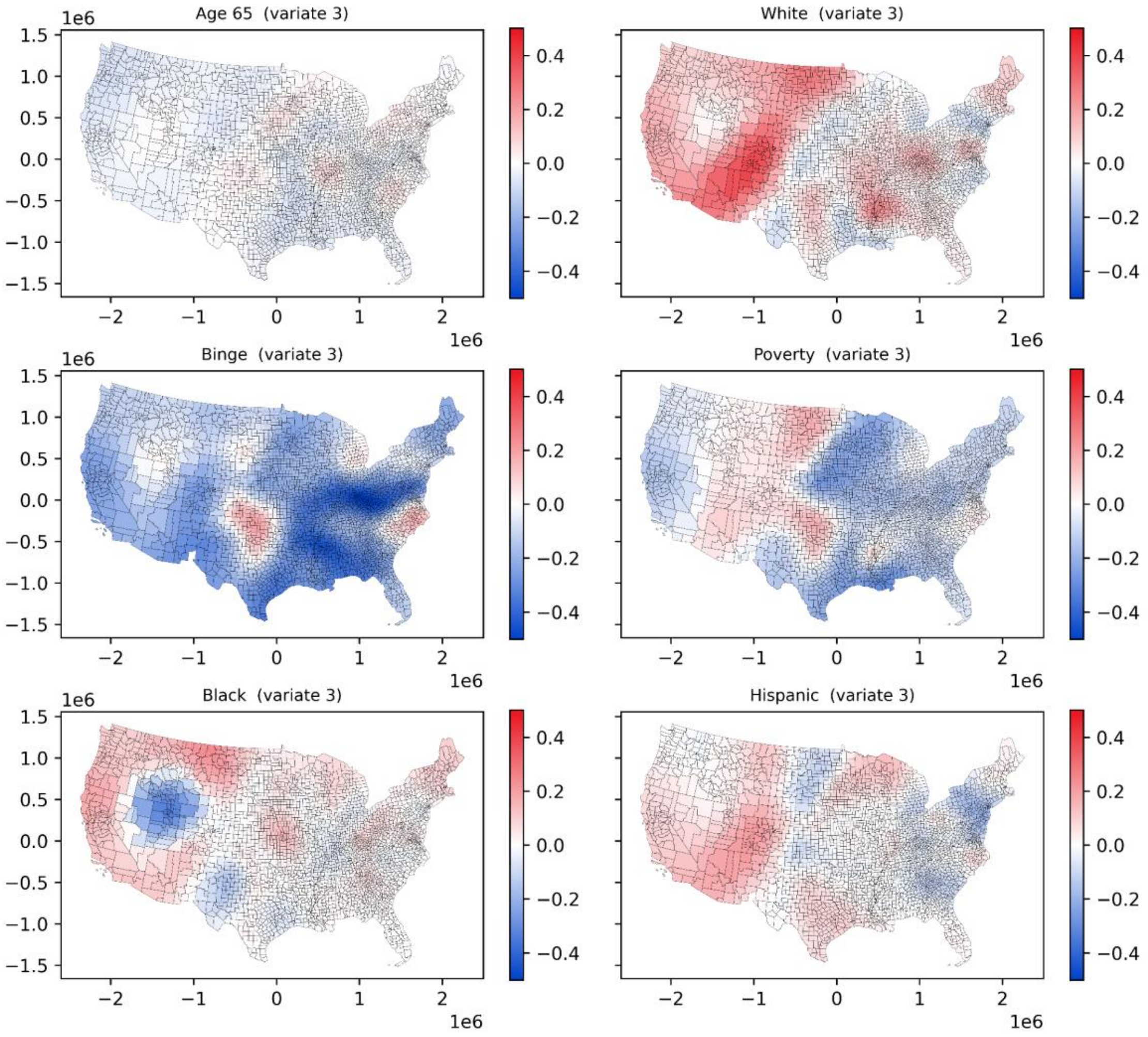

Figure C3. Local variable loadings of SDOH in canonical variate 3

## Appendix D

All experiments were conducted on Google Colab using a virtualized Linux environment. The runtime was equipped with an Intel Xeon CPU operating at 2.20 GHz, featuring 2 logical CPUs under a KVM-based full virtualization setup. The system architecture was x86_64 (64-bit) with little-endian byte order. The available system memory was 12 GB RAM.

Table D1. Computational time and resource of GWCCA for the study datasets.

| Datasets | Time consumption | RAM consumption |
| --- | --- | --- |
| Synthetic dataset I | 12 mins | 500 MB |
| Synthetic dataset II | 15 mins | 600 MB |
| Case study | 25 mins | 1.5 GB |